\documentclass[11pt]{article}

\usepackage{geometry}

\usepackage{graphicx,amsfonts,bm,epsfig,color,latexsym}
\usepackage{amsmath}
\usepackage{amsfonts}
\usepackage{amssymb}
\usepackage{epstopdf}
\usepackage{float}
\definecolor{background-color}{gray}{0.98}

\newcommand{\bea}{\begin{eqnarray}}
\newcommand{\eea}{\end{eqnarray}}
\newcommand{\bes}{\begin{subequations}}
	\newcommand{\ees}{\end{subequations}}
\newtheorem{dfn}{Definition}[section]

\newtheorem{theorem}{Theorem}[section]

\newcommand{\bd}{\begin{document}}
	\newcommand{\ed}{\end{document}}
\newcommand{\bc}{\begin{center}}
	\newcommand{\ec}{\end{center}}
\newcommand{\bfr}{\begin{flushright}}
	\newcommand{\efr}{\end{flushright}}
\newcommand{\lt}{\left}
\newcommand{\rt}{\right}
\newcommand{\vs}{\vspace}
\newcommand{\hs}{\hspace}
\newcommand{\beq}{\begin{equation}}
\newcommand{\eeq}{\end{equation}}
\newcommand{\lb}{\linebreak}
\newcommand{\pb}{\pagebreak}
\newcommand{\mb}{\makebox}
\newcommand{\fb}{\framebox}
\newcommand{\mc}{\multicolumn}
\newcommand{\ben}{\begin{enumerate}}
	\newcommand{\een}{\end{enumerate}}
\newcommand{\bit}{\begin{itemize}}
	\newcommand{\eit}{\end{itemize}}
\newcommand{\oln}{\overline}
\newcommand{\un}{\underline}
\newcommand{\lefq}{\lefteqn}
\newcommand{\ba}{\begin{array}}
	\newcommand{\ea}{\end{array}}
\newcommand{\beqa}{\begin{eqnarray}}
\newcommand{\eeqa}{\end{eqnarray}}
\newcommand{\beqas}{\begin{eqnarray*}}
	\newcommand{\eeqas}{\end{eqnarray*}}
\newcommand{\bfg}{\begin{figure}}
	\newcommand{\efg}{\end{figure}}
\newcommand{\bds}{\begin{displaymath}}
\newcommand{\eds}{\end{displaymath}}
\newcommand{\btb}{\begin{tabbing}}
	\newcommand{\etb}{\end{tabbing}}
\newcommand{\para}{\parallel}
\newcommand{\pad}{\partial}
\newcommand{\nn}{\nonumber}
\newcommand{\la}{\leftarrow}
\newcommand{\ra}{\rightarrow}
\newcommand{\lgla}{\longleftarrow}
\newcommand{\lgra}{\longrightarrow}
\newcommand{\La}{\Leftarrow}\newcommand{\Ra}{\Rightarrow}
\newcommand{\Lra}{\Leftrightarrow}
\newcommand{\Lgla}{\Longleftarrow}
\newcommand{\Lgra}{\Longrightarrow}
\newcommand{\lan}{\langle}
\newcommand{\ran}{\rangle}
\renewcommand{\a}{\alpha}
\renewcommand{\b}{\beta}
\newcommand{\g}{\gamma}
\newcommand{\G}{\Gamma}
\renewcommand{\d}{\delta}
\newcommand{\eps}{\epsilon}
\newcommand{\Th}{\Theta}
\newcommand{\s}{\sigma}
\newcommand{\lam}{\lambda}
\newcommand{\D}{\Delta}
\newcommand{\ds}{\displaystyle}
\newcommand{\vare}{E}
\newcommand{\pr}{\prime}
\newcommand{\ro}{\rho}
\newcommand{\nab}{\nabla}
\newcommand{\m}{\mu}
\newcommand{\n}{\nu}
\newcommand{\Sg}{\Sigma}
\newcommand{\p}{\pi}
\newcommand{\R}{I\!\!R}
\newcommand{\om}{\omega}
\newcommand{\Om}{\Omega}
\newcommand{\ovra}{\overrightarrow}
\newcommand{\ze}{\zeta}
\newcommand{\vart}{\vartheta}
\newcommand{\tri}{\triangle}
\newcommand{\f}{\frac}
\newcommand{\iny}{\infty}
\newcommand{\pro}{\propto}
\renewcommand{\arraystretch}{1.25}

\title{Properties of R\'enyi complexity ratio of quantum states for central potential}
\author{Debraj Nath\thanks{Department of Mathematics, Vivekananda College, Thakurpukur, Kolkata-700063, India. Email: debrajn@gmail.com}}
\begin{document}
	
	\maketitle
	
	% ORCID Piu Ghosh 0000-0002-3568-783X
	% ORCID Debraj Nath 0000-0001-9937-7032			
	\begin{abstract}
	R\'enyi complexity ratio of two density functions is introduced for three and multidimensional quantum systems. Localization property of several density functions are defined and five theorems about near continuous property of R\'enyi complexity ratio are proved by Lebesgue measure. Some properties of R\'enyi complexity ratio are demonstrated and investigated for different quantum systems. Exact analytical forms of R\'enyi entropy, R\'enyi complexity ratio, statistical complexities based on R\'enyi entropy for integral order have been presented for solutions of pseudoharmonic and a family of isospectral potentials. Some properties of R\'enyi complexity ratio are verified for some diatomic molecules (CO, NO, N$_2$, CH, H$_2$, and ScH) and for some other quantum systems. \\\\%It is an extension of the generalized R\'enyi complexity.
\end{abstract}
\bibliographystyle{apsrev}

\renewcommand{\baselinestretch}{1.5}
\normalsize
\section{Introduction}
Information entropy and statistical complexity is a growing interesting subject for studying the behavior of atomic structure in physics and quantum chemistry. Specially the Shannon entropy \cite{Shannon} and the R\'enyi entropy \cite{Renyi} are more useful measurements for entropic uncertainty relations \cite{GEUR,BBM}, in atomic system and statistical thermodynamics \cite{Renyi.stat,Renyi.stat2}. The R\'enyi entropy is important in quantum chemistry \cite{stat.complexity}, mathematical physics \cite{Renyi.jsd}, quantum information \& quantum computation \cite{Renyi.infor}, statistical mechanics \cite{Renyi.statis}, image processing \cite{Renyi.image}, computer science \cite{Renyi.comp} and different fields of science. It is used as a generalization of Shannon entropy. It has many applications in quantum information and some interesting physical measurement can de defined by it \cite{pipek,BBM,Renyi.appl.nagy,Renyi.appl,Renyi.appl2,s.liu1997,R.G.Parr.BOOK}. In momentum space, it is defined for shell structure of atoms \cite{Renyi.nagi.mom}.	   

A useful and important statistical complexity is L\'opez-Ruiz-Mancini-Calbet (LMC) complexity \cite{lmc,disequilibrium2,lmc.plasito.continuous,lmc.continuous}. It is defined, as a product Shannon entropy and disequilibrium \cite{Hall}. Another simple measure of complexity is Shiner, Davison, Landsber (SDL), which is a product of order and disorder of a quantum state \cite{SDL}. The LMC is modified and known as shape LMC \cite{disequilibrium2,lmc.continuous,Yamano.JMP2004,lmc.biophys,lmc.shape.jca}, which is a product of power of Shannon entropy and disequilibrium. The LMC and shape LMC complexities have been applied in different fields of science \cite{Yamano.JMP2004, lmc.appl}. Moreover, shape LMC is modified, so-called shape R\'enyi complexity (SRC) \cite{jc.anguloCPL,a.nagy.IRP}, where Shannon entropy is replaced by R\'enyi entropy. Again the shape R\'enyi complexity is modified, so-called generalized R\'enyi complexity (GRC) \cite{a.nagy.jmp,a.nagy.chaos, jc.angulo.jsd.epjd,jsd.pla2016, DN.ijmpa, DN.IJQC.Renyi}, where disequilibrium is replaced by inverse of R\'enyi entropic power. The SRC is a one parameter family of complexity measure, whereas GRC is a two parameters family complexity. The LMC, SRC and GRC have several properties and applications in physics, mainly in quantum chemistry for atomic structure. But there is an example to prove the near continuous property of LMC, SRC, GRC and there is no analytical prove for arbitrary density functions \cite{lmc.continuous, a.nagy.jmp,nagy.j.stat.mec}. 

Different types of complexities such as Fisher-Shannon \cite{FS} for ionization processes ans Fisher-R\'enyi for atomic density function \cite{Renyi.nagy.FR} are investigated in some literatures. All such complexities are defined for a single density function. In addition, some conditional or relative information, such as (i) relative Shannon \cite{relative.S}, (ii) relative Fisher \cite{relative.F}, (iii) relative R\'enyi \cite{Renyi.nagy.relative} and (iv) relative Tsallis \cite{relative.Ts} have been defined between two density functions. The relative Shannon \cite{relative.S}, relative R\'enyi \cite{Renyi.nagy.relative} are used in atomic system. The relative Fisher has been used for central potential. A relative LMC-type complexity is defined for atoms \cite{relative.lmc,relative.lmc.rc} and a generalized relative complexity \cite{relative.lmc.rc} is defined for Dicke model \cite{Renyi.nagy.relative, nagy.j.stat.mec} by the definitions of relative Shannon and relative R\'enyi entropies. Recently complexity ratio has been introduced in position and momentum spaces for radial pseudoharmonic oscillator potential \cite{pshop9}. With respect to generalized quantum similarity index \cite{selfsimilarity}, we already shown that, wave functions of some diatomic molecules are same for pseudoharmonic oscillator which match with a family isospectral potentials in 3D \cite{DN.ijqc.qsi}.

In this paper, the R\'enti complexity ratio (RCR) of two density functions will be introduced. The aim of this paper is to find the relation between RCR and GRC. To this aim, some definitions of localization property of several density functions will be defined and some theorems of near continuous property of RCR in  different dimensions will be proved by Lebesgue measure. But the main focus of this paper is to explore the R\'enyi complexity ratio in three dimensional quantum systems for central potential. 

The RCR is an extension of GRC, which is an interesting field of quantum chemistry. The definition of RCR can be applied directly to GRC, SRC and LMC for two identical density functions, as a particular case of RCR. All properties of RCR will be verified to the solutions of pseudoharmonic oscillator \cite{pshopEx1.2,pshop.E.Kasap,Flugge,pshopEx2,pshop2.R.Sever.theo.chem,pshop3.R.Sever.jmc2007,pshop4,pshop5.R.Sever.jmc2012,pshop6.kds.oyewumi.jmc2012,pshop7.R.Sever.jmc2012.1973,pshop8.H.Hassanbadi,pshop9,pshop10.F.Chand.Pramana,DN.ijqc.pho} and a family of isospectral potentials \cite{SUSY}. Moreover, the effect of isospectral parameter $\lam$ on R\'enyi entropy, RCR, GRC and SRC will be examined. An interesting limiting case $\left(|\lam|\rightarrow\infty\right)$, which will be considered for the measures of RCR, GRC and SRC. Moreove we will examine the effect of $D_e$ (dissociation energy of diatomic molecules), $r_e$ (equilibrium intermolecular separation), $m_{\mu}$ (reduced mass of diatomic molecules) and $\lam$ (isospectral parameter) on R\'enyi entropy, RCR, GRC and SRC.

This paper is divided into four sections. They are organized as follows. In Sec. \ref{section.GRC}, localization property of several density functions will be introduced. Then some known results of R\'enyi entropy will rewritten. In Sec. \ref{section.main}, main idea for R\'enyi complexity ratio which is an extension of generalized R\'enyi complexity will be explained. Next some theorems and properties of RCR will be demonstrated and investigated. In Sec. \ref{section.appl}, the exact forms of R\'enyi entropy, RCR, GRC and SRC of the solutions of pseudoharmonic oscillator and a family of isospectral potentials will be presented. In Sec. \ref{results} R\'enyi entropy, RCR, GRC and SRC will be calculated numerically for rational orders for 19 diatomic molecules and some other quantum systems. Beside this example, this method will be worked for solutions of any other potentials. Finally, some conclusion will be given in Sec. \ref{conclusion}.

\section{Preliminaries about density functions and R\'enyi entropy}\label{section.GRC}
\subsection{Localization property of density functions}
All density functions are bounded in a region $\mathbb{R}^g=[0,\infty)\times[0,\pi]\times[0,2\pi]$ for a three dimensional spherical coordinates and the joint density function $\rho:\mathbb{R}^g\rightarrow[0,1]$ can be expressed as a product of three independent density functions $f(r)$, $g(\theta)$ and $h(\phi)$. For a central potential one of them, $h(\phi)=\f{1}{2\pi}$, $\phi\in[0,2\pi]$ is a uniform density function and there exists a $d>0$ such that 
\beq\label{int.r}
\ds\int_{0}^{\infty} f(r) r^2dr=1=\ds\int_{0}^{d} f(r) r^2dr,~\ds\int_{d}^{\infty} f(r) r^2dr=0.
\eeq
Integrations in Eq.(\ref{int.r}) are considered as a measure of Lebesgue integration. One can define a partition of the interval $[0,d]$ by disjoint sets, such that \cite{pipek,DN.IJQC.Renyi} $[0,d]=\ds\cup_{i=1}^{N} \Om_i$ and
\beq
\ba{l}
\sum\limits_{i=1}^N p_i=1,~0\le p_i\le 1,~\mbox{where~}
p_i=\ds\int_{\Om_i}f(r)r^2 dr.
\ea
\eeq 
Under the region $\Om=\cup_{i}\Om_i$, where $p_i\ne 0$, one can obtain 
\beq
\ds\int_{0}^{\infty} f(r) r^2dr=1=\ds\int_{\Om} f(r) r^2dr.
\eeq
Then $\Om$ is the actual effective domain of the density function $f(r)$. In this paper, the Lebesgue measure of $\Om$ is written by $\mathcal{L}(\Om)$. Several effective domains with respect to radial distance $r$, polar angle $\theta$, azimuthal angle $\phi$ and ${\bf r}$ of their corresponding density functions can be defined by Lebesgue measure

\noindent \begin{dfn} \label{def.eff.r} 
	(Effective domain with respect to $r$ \cite{DN.IJQC.Renyi}). A region $\Om_r\subset[0,\infty)$ is called the effective domain of the density function $f(r)$, if $\ds\int_{\Om_r} f(r) r^2dr=1$ and there exists no $\Om_1\subset[0,\infty)$ such that $\mathcal{L}(\Om_1)<\mathcal{L}(\Om_r)$, where $\ds\int_{\Om_1} f(r) r^2dr=1$. On the other hand $\Om_r$ is called the effective domain of $f(r)$, if $\ds\int_{\Om_r} f(r) r^2dr=1$ and $\mathcal{L}(\Om_r)\le \mathcal{L}(\Om_1)$ for any $\Om_1\in\left\{\Om\subset[0,\infty): \ds\int_{\Om} f(r) r^2dr=1\right\}$.
	\end{dfn}

\noindent \begin{dfn}\label{def.loc.r} 
	(Localization with respect to $r$ \cite{DN.IJQC.Renyi}). Let $f_1(r)$ and $f_2(r)$ be two radial density functions, $\Om_r^1$ and $\Om_r^2$ be the effective domains of $f_1$ and $f_2$ respectively with respect to $r$. Then $f_1$ is called localized than $f_2$ with respect to $r$ if $\mathcal{L}(\Om_r^1)< \mathcal{L}(\Om_r^2)$. 
\end{dfn}

\noindent \begin{dfn} \label{def.eff.th} 
	(Effective domain with respect to $\theta$ \cite{DN.IJQC.Renyi}). A region $\Om_{\theta}\subset[0,\pi]$ is called the effective domain of the density function $f(\theta)$, if $\ds\int_{\Om_{\theta}} f(\theta)\sin\theta\,d\theta=1$ and there exists no $\Om_1\subset[0,\pi]$ such that $\mathcal{L}(\Om_1)<\mathcal{L}(\Om_{\theta})$, where $\ds\int_{\Om_1} f(\theta)\sin\theta\,d\theta=1$. Alternatively $\Om_{\theta}$ is called the effective domain of $f(\theta)$, if $\ds\int_{\Om_{\theta}} f(\theta)\sin\theta\,d\theta=1$ and $\mathcal{L}(\Om_{\theta})\le \mathcal{L}(\Om_1)$ for any $\Om_1\in\left\{\Om\subset[0,\pi]: \ds\int_{\Om} f(\theta)\sin\theta\,d\theta=1\right\}$.
\end{dfn}.	
\noindent \begin{dfn} \label{def.loc.th}
	(Localization with respect to $\theta$ \cite{DN.IJQC.Renyi}). Let $f_1(\theta)$ and $f_2(\theta)$ be two rotational density functions of the polar angle $\theta$, $\Om_{\theta}^1$ and $\Om_{\theta}^2$ be the effective domains of $f_1$ and $f_2$ respectively with respect to $\theta$. Then $f_1$ is called localized than $f_2$ with respect to $\theta$ if $\mathcal{L}(\Om_{\theta}^1)<\mathcal{L}(\Om_{\theta}^2)$.
\end{dfn}

\noindent \begin{dfn} \label{def.eff.ph} 
	(Effective domain with respect to $\phi$). A region $\Om_{\phi}\subset[0,2\pi]$ is called the effective domain of the density function $f(\phi)$, if $\ds\int_{\Om_{\phi}} f(\phi)d\phi=1$ and there exists no $\Om_1\subset[0,2\pi]$ such that $\mathcal{L}(\Om_1)<\mathcal{L}(\Om_{\phi})$, where $\ds\int_{\Om_1} f(\phi)d\phi=1$. Alternatively effective domain $\Om$ of $f(\phi)$ can be defined as, $\ds\int_{\Om_{\phi}} f(\phi)d\phi=1$ and $\mathcal{L}(\Om_{\phi})\le \mathcal{L}(\Om_1)$ for any $\Om_1\in\left\{\Om\subset[0,2\pi]: \ds\int_{\Om} f(\phi)d\phi=1\right\}$.
\end{dfn}.	
\noindent \begin{dfn} \label{def.loc.ph}
	(Localization with respect to $\phi$). Let $f_1(\phi)$ and $f_2(\phi)$ be two density functions of the azimuthal angle $\phi$, $\Om_{\phi}^1$ and $\Om_{\phi}^2$ be the effective domain of $f_1$ and $f_2$ respectively with respect to $\phi$. Then $f_1$ is called localized than $f_2$ with respect to $\phi$ if $\mathcal{L}(\Om_{\phi}^1)<\mathcal{L}(\Om_{\phi}^2)$.
\end{dfn}
\noindent \begin{dfn}\label{def.eff.vecr}
	(Effective domain with respect to ${\bf r}$). Let a joint density function $\rho({\bf r})=\rho_{r}(r)\rho_{\theta}(\theta)\rho_{\phi}(\phi)$ be defined in $\mathbb{R}^g$. If there exist $\Om_r\subset[0,\infty)$, $\Om_{\theta}\subset [0,\pi]$, $\Om_{\phi}\subset [0,2\pi]$ such that $\ds\int_{\Om_r\times\Om_{\theta}\times\Om_{\phi}}\rho({\bf r})d{\bf r}=1$, $\ds\int_{\Om_r}\rho_r(r)r^2\,dr=\ds\int_{\Om_{\theta}}\rho_{\theta}(\theta)\sin\theta\,d\theta=\ds\int_{\Om_{\phi}}\rho_{\phi}(\phi)d\phi=1$, and  $\mathcal{L}(\Om_r)\le\mathcal{L}(\Om_1)$, $\mathcal{L}(\Om_{\theta})\le\mathcal{L}(\Om_2)$, $\mathcal{L}(\Om_{\phi})\le\mathcal{L}(\Om_3)$ for all $\Om_1\in\left\{\Om\subset[0,\infty):\right.$ $\left. \ds\int_{\Om} \rho_{r}(r)r^2\, dr\right\}$; $\Om_2\in\left\{\Om\subset[0,\pi]:\ds\int_{\Om} \rho_{\theta}(\theta) \sin\theta\, d\theta=1\right\}$; and $\Om_3\in\left\{\Om\subset[0,2\pi]:\ds\int_{\Om} \rho_{\phi}(\phi) d\phi=1\right\}$, then the region $\Om_r\times\Om_{\theta}\times\Om_{\phi}$ is called the effective domain of $\rho({\bf r})$. 
	\end{dfn}

It is to be noted that, the effective domain is an area or a volume  of a two or three dimensional quantum system. One can define the effective domain of a $D$ dimensional quantum system. It is difficult to find the effective domain of a multi-dimensional non-separable quantum state. But one can define the localization property, for a multi-dimensional non-separable quantum system with known density functions. In this paper, three dimensional quantum system for central potential is considered.

\subsection{R\'enyi entropy, R\'enyi length and R\'enyi volume}
Let a density function $\rho:M^D\rightarrow[0,1]$ is defined on a $D$-dimensional space $M^D\subset\mathbb{R}^D$. The one-parameter (order $\a$) R\'enyi entropy of $\rho$ is defined by \cite{Renyi}
\beq\label{Renyi.con}
\mathcal{R}_{\rho}^{(\a)}=\f{1}{1-\a}\ln\left[I^{(\a)}\right],~ \a>0,~ \ne 1,
\eeq
where $I^{(\a)}$ is the entropic moments of the density function $\rho$ and is defined by %\cite{jkord}
\beq
I^{(\a)}=\int_{\Om^D} \rho^{\a}({\bf r}) d{\bf r},
\eeq
and $\Om^D\subset M^D$ is the effective domain of $\rho$. If the effective domain of $\rho$ exists, then R\'enyi entropy (\ref{Renyi.con}) can be written as
\beq\label{Renyi.dis}
\ba{l}
\mathcal{R}_{\rho}^{(\a)}=\f{1}{1-\a}\ln\sum\limits_{i} p_i^{\a}\mathcal{L}(\Om_i),~ \a>0,~ \ne 1,
\ea
\eeq
where
\beq
\ba{l}
\ds\sum\limits_{i} p_i\mathcal{L}(\Om_i)=1,~0\le p_i\le 1,~\mathcal{L}(\Om_i)\ge 0,\\
\rho({\bf r})=\ds\sum\limits_{i}p_i\,\chi_{\Om_i}({\bf r}),\\
\ba{ll}
\chi_{\Om_i}({\bf r})&=1,{\bf r}\in\Om_i,\\
&=0,{\bf r}\notin\Om_i,
\ea
\ea 
\eeq 
and $\Om_{\rho}^D=\cup_{i}\Om_i$. The sum (\ref{Renyi.dis}) is a good approximation of (\ref{Renyi.con}), if $\mathcal{L}(\Om_i)\rightarrow 0$. For a discrete distribution R\'enyi entropy is written by $\mathcal{R}_{\rho}^{(\a)}=\f{1}{1-\a}\ln\sum\limits_{i} p_i^{\a}$. Also it can be obtained form (\ref{Renyi.dis}), if $\mathcal{L}(\Om_i)=1$ and $\ds\sum\limits_{i}\mathcal{L}(\Om_i)=\mathcal{O}(\Om_{\rho}^D)=$ \emph{order of the set $\Om_{\rho}^D$ or size of the distributions}. The R\'enyi entropy defined in (\ref{Renyi.dis}) of a finite discrete distribution is always positive, but (\ref{Renyi.con}) may be negative \cite{Renyi.appl.nagy} for sufficiently large variations. It is a non-increasing function of order $\a$. The Tsallis entropy is an another important family of generalized entropy and it is defined by \cite{tsallis}
\beq
\mathcal{T}_{\rho}^{(\a)}=\f{1}{\a-1}\left(1-I^{(\a)}\right),~ \a>0,~ \ne 1.
\eeq
A relation between R\'enyi and Tsallis entropies is 
\beq
\mathcal{T}_{\rho}^{(\a)}=\f{1}{1-\a}\left(e^{(1-\a)\mathcal{R}_{\rho}^{(\a)}}-1\right),
\eeq
or
\beq
\mathcal{R}_{\rho}^{(\a)}=\f{1}{1-\a}\ln\left[1+(1-\a)\mathcal{T}_{\rho}^{(\a)}\right].
\eeq
If $\rho$ is a delta function, then both are equal to zero. In the limiting case $(\a\rightarrow1)$, they reduce to the Shannon entropy $\mathcal{S}_{\rho}$ and it is defined by \cite{Shannon}
\beq
\mathcal{S}_{\rho}=-\ds\int\rho({\bf r})\ln\rho({\bf r})\,d{\bf r}.
\eeq 
%If $D=1,2$ and $3$, then $e^{\mathcal{R}_{\rho}^{(\a)}}$ has a dimension of length, area and volume respectively \cite{Hall}. 
R\'enyi volume and length are denoted by $\mathcal{V}^{(\a)}_{\rho}$ and ($\mathcal{L}^{(\a)}_{(\rho)}$) respectively and for $D=3$ a relation between them is defined by \cite{pshop9,impetus}
\beq
\mathcal{L}^{(\a)}_{\rho}=\ds\left(\f{3}{4\pi}\mathcal{V}_{\rho}^{(\a)}\right)^{\f{1}{3}}=\ds\left(\f{3}{4\pi}\right)^{\ds\f{1}{3}}e^{\f{1}{3}\mathcal{R}_{\rho}^{(\a)}},~\a>0,\ne 1.
\eeq
For a special case $\a=0$ and for continuous distribution the R\'enyi entropy $\mathcal{R}_{\rho}^{(0)}$ is defined by $\ln\mathcal{L}(\Om_{\rho}^D)$. %If $D=1,2$ and $3$, then $\mathcal{R}_{\rho}^{(0)}$ is a function of length, area and volume respectively. 
For discrete random variable $\mathcal{R}_{\rho}^{(0)}$ is equal to \cite{Renyi.stat,Renyi.stat2} $\ln \mathcal{O}(\Om_{\rho}^D)$. 

Moreover, for $\a=\f{4}{3}, \f{5}{3}, 2$, R\'enyi entropy is related with some physical quantities, such as Thomas-Fermi kinetic energy, the Dirac exchange energy and electron density. It has many applications in density functional theory for atoms and molecules \cite{s.liu1997,R.G.Parr.BOOK,Renyi.appl.nagy}. The average density is called the disequilibrium and it has dimension of inverse volume. It is inversely proportional to the R\'enyi volume of order 2 and it is defined by \cite{Onicescu,selfsimilarity,disequilibrium2,Hall}, $\mathcal{D}_{\rho}=e^{-\mathcal{R}_{\rho}^{(2)}}$. On the other hand $\mathcal{R}_{\rho}^{(1)}-\mathcal{R}_{\rho}^{(1)}$ is defined the structural entropy \cite{pipek} of $\rho$. An another special case is \cite{debnath} 
\beq
\mathcal{R}_{\rho}^{(\a)}\rightarrow-\ln\parallel\rho\parallel_{\infty},~\mbox{if~}\a\rightarrow\infty,
\eeq  
where $\parallel\rho\parallel_{\infty}=\sup\limits_{{\bf r}}\rho({\bf r})$. The R\'enyi entropy satisfies several properties \cite{pipek,BBM,Renyi.appl,Renyi.appl.nagy,Renyi.appl2} and some important inequalities which are relevant to this work \cite{Renyi.stat} are written as follows
\beq\label{Renyi.ineq}
\ba{ll}
\f{\partial}{\partial\a}\mathcal{R}_{\rho}^{(\a)}&\le 0,\\
\f{\partial}{\partial \a}\left(\f{\a-1}{\a}\mathcal{R}_{\rho}^{(\a)}\right)&\ge 0,\\
\mathcal{R}_{\rho}^{(1)}&\ge 2\mathcal{R}_{\rho}^{(2)}-\mathcal{R}_{\rho}^{(3)}.
\ea
\eeq
\section{R\'enyi complexity ratio}\label{section.main}
R\'enyi complexity ratio of two density functions $f$ and $g$ of order $(\a,\b)$ is defined by 
\beq\label{newcomplexity}
C^{(\a,\b)}_{(f,g)}=e^{\mathcal{R}^{(\a)}_{f}-\mathcal{R}^{(\b)}_{g}}.
\eeq 
\subsection{Simple general properties of RCR}
The R\'enyi complexity ratio of two density functions $f$ and $g$ satisfies several properties. They are as follows:\\
%\begin{itemize}
	\noindent $\bullet$ (i) $C^{(\a,\b)}_{(f,f)}$ reduces to GRC \cite{a.nagy.IRP,a.nagy.jmp,a.nagy.chaos,jc.angulo.jsd.epjd,jsd.pla2016,DN.ijmpa} of $f$ with order $(\a,\b)$.\\
	\noindent $\bullet$ (ii) $C^{(\a,\b)}_{(f,g)}C^{(\a,\b)}_{(g,f)}=C_{(f,f)}^{(\a,\b)}C_{(g,g)}^{(\a,\b)}$.\\
	\noindent $\bullet$ (iii) $C^{(\a,\b)}_{(f,g)}C^{(\b,\a)}_{(g,f)}=1$,~ $C^{(\a,\b)}_{(f,f)}C^{(\b,\a)}_{(f,f)}=1$.\\
	\noindent $\bullet$ (iv) $C^{(\a,\a)}_{(f,f)}=1$.\\
	\noindent $\bullet$ (v) $C^{(\a,\b)}_{(f,g)}$ is a non-increasing function of $\a$ for fixed $\b$ and $g$. It is an increasing function of $\b$ for fixed $\a$ and $f$.\\
	\subsection{Majorization effect on RCR}
	%%%%%%%%%%%%%%%%%%%
	\begin{dfn}
		(Majorization for FDD \cite{majorization}) Let $\rho_j=(p_1^j,p_2^j,\dots,p_n^j)$, $j=1,2$ be two finite discrete distributions. Then $\rho_1$ majorizes $\rho_2$ $(\rho_1\succ\rho_2)$, if $\sum\limits_{i=1}^k p_i^{\downarrow 1}\ge  	\sum\limits_{i=1}^k p_i^{\downarrow 2},~1\le k<n$, and $\sum\limits_{i=1}^n p_i^{\downarrow 1}=	\sum\limits_{i=1}^n p_i^{\downarrow 2}=1$, where $p_1^{\downarrow j}=\max\left\{p_i^j:i=1,2,\dots,n\right\},~p_n^{\downarrow j}=\min\left\{p_i^j:i=1,2,\dots,n\right\},~p_1^{\downarrow j}\ge p_2^{\downarrow j}\ge ...\ge p_n^{\downarrow j}$.
	\end{dfn}
	%%%%%%%%%%%%%%%%%%%%
	\begin{dfn}
		(Majorization for CD \cite{joe})
		Let $\rho_1:M^D\rightarrow[0,1]$ and $\rho_2:M^D\rightarrow[0,1]$ be two density functions. Then $\rho_1$ majorizes $\rho_2$ $\left(\rho_1\succ \rho_2\right)$, if $\ds\int\left[\rho_1({\bf r})-r_0\right]^+d{\bf r}\ge\ds\int\left[\rho_2({\bf r})-r_0\right]^+d{\bf r}$, holds, for all $r_0\ge 0$, where $\left[y\right]^+=\max\left\{y,0\right\}$. 
		%$d\mu$ is the measure defined on $I$. 
	\end{dfn}
%%%%%%%%%%%%%%%%%%%%%%%%%%%%%
The entropic moment $I^{(\a)}$ is a concave functional if $0<\a<1$ or convex functional if $\a>1$ of density functions \cite{majorization,joe,majorization.open,majorization.BOOK,puchala}. Therefore, 
\beq
\rho_1\succ \rho_2\implies\left\{\ba{ll}I^{(\a)}_{\rho_1}\le I^{(\a)}_{\rho_2}&\mbox{if~}0<\a<1\\ I^{(\a)}_{\rho_1}\ge I^{(\a)}_{\rho_2}&\mbox{if~}\a>1\ea\right. .
\eeq 
and R\'enyi entropy, $\mathcal{R}_{\rho}^{(\a)}$ is a concave functional of $\rho$, for any $\a>0$ and hence one can write 
\beq
\rho_1\succ \rho_2\implies\mathcal{R}_{\rho_1}^{(\a)}\le \mathcal{R}_{\rho_2}^{(\a)},~\a>0.
\eeq
%%%%%%%%%%%%%%%%%%%%%%%%%%%%%%%%%%%%%
	\noindent $\bullet$ (vi) It is well known that, R\'enyi entropy $\mathcal{R}_f^{(\a)}$ is a non-increasing function of $\a$ and one can write \cite{DN.IJQC.Renyi} $\mathcal{R}_f^{(\a)}>\mathcal{R}_g^{(\a)}$ if $f$ is widely spread and $g$ is narrowly confined on a domain and for two discrete distributions \cite{majorization} $\mathcal{R}_f^{(\a)}>\mathcal{R}_g^{(\a)}$ if $f<g$. Hence, one can write nature of $C_{(f,g)}^{(\a,\b)}$ which satisfies some inequalities as follows  
	\beq\label{majo.RCR}
	C^{(\a,\b)}_{(f,g)}\left\{\ba{ll}
	\le 1, & \a\ge \b,~ f \succ g\\
	\cdots, & \a> \b,~f\prec g \\
	\ge 1, & \a\le\b,~ f\prec g \\
	\cdots,    & \a< \b,~ f\succ g 
			\ea \right. .
	\eeq 
	Now one can find lower and upper bound of $C_{(f,g)}^{(\a,\b)}$ using majorization effect (\ref{majo.RCR}).\\
\noindent $\bullet$ (vii) Upper bound of $C^{(\a,\b)}_{(f,g)}=1$, exists, if $f\succ g$ and $\a>\b$.\\	
\noindent $\bullet$ (viii) Lower bound of $C^{(\a,\b)}_{(f,g)}=1$ exists, if $f\prec g$ and $\a<\b$.\\
	\noindent $\bullet$ (ix) The RCR is a postive define functional of density functions, therefore, in general lower bound of $C^{(\a,\b)}_{(f,g)}$ is zero and the upper bound may be defined by R\'enyi entropic bound. It is already known that, $\mathcal{R}_g^{(\b)}\ge \mathcal{R}_g^{(\infty)}$, for $\b>0$. Now, if two density functions $f({\bf r})$ and $g({\bf r})$  are defined in a $D$-dimensional central potential, then one can write \cite{jsd.jmp.Renyi.bound}
	\beq\label{RCR.bound}
	C_{(f,g)}^{(\a,\b)}\le\parallel g\parallel_{\infty}\ds e^{\mathcal{B}_D(\a)}\left(\f{\left\langle r^2\right\rangle_f}{D}\right)^{\f{D}{2}},
	\eeq 
	where
	\beq
	\parallel g\parallel_{\infty}=\sup\limits_{{\bf r}} g({\bf r}), \left\langle r^2\right\rangle_f=\int_{M^D}f({\bf r})r^2 d{\bf r},
	\eeq 
	and
%	\begin{widetext}
	\beq
	\ba{lll}
	\mathcal{B}_D(\a)&=\f{D}{2}\log\left[\f{\pi((2+D)\a-D)}{1-\a}\right]-\f{\a}{1-\a}\log\left[\f{(2+D)\a-D}{2\a}\right]-\log\left[\f{\G(\f{\a}{1-\a})}{\G(\f{(2+D)\a-D}{2(1-\a)})}\right],&\f{D}{D+2}<\a<1\\
	&=\f{D}{2}\log(2\pi e),&\a=1\\
	&=\f{D}{2}\log\left[\f{\pi((2+D)\a-D)}{\a-1}\right]+\f{\a}{\a-1}\log\left[\f{(2+D)\a-D}{2\a}\right]+\log\left[\f{\G(\f{\a}{\a-1})}{\G(\f{(2+D)\a-D}{2(\a-1)})}\right],&\a>1
	\ea.
	\eeq
%	\end{widetext} 
%	{\bf add upper bounds  ***********.....}\\ 
	So the upper bound depends on $f$, $g$ and $(\a,\b)$. Now using the inequalities (\ref{Renyi.ineq}), one can improve the inequality (\ref{RCR.bound}) as
	\beq
	\ba{ll}
	C_{(f,g)}^{(\a,\b)}%&\le\min\left\{\mathcal{G}_g(\b),\parallel g\parallel_{\infty}\right\}e^{\left[\mathcal{B}_D(\a)+\f{D}{2}\ln\left(\f{\left\langle r^2\right\rangle_f}{D}\right)\right]},\\
	&\le\min\left\{\mathcal{G}_g(\b),\parallel g\parallel_{\infty}\right\}\left(\f{\left\langle r^2\right\rangle_f}{D}\right)^{\f{D}{2}}e^{\mathcal{B}_D(\a)},
	\ea 
	\eeq 
	where
	\beq
	\ba{lll}
	\mathcal{G}_g(\b)&=\left(\parallel g\parallel_{\infty}\right)^{\f{-\b}{1-\b}},&\b<1\\
	&=\f{\left(\mathcal{D}_g\right)^2}{\sqrt{\mathcal{R}_g^{(3)}}},&\b=1\\
	&=\parallel g\parallel_{\infty},&\b>1
	\ea.
	\eeq 
	
\noindent $\bullet$ (x) 
\beq\label{Cfg.xiia}
C^{(\a,\b)}_{(f,g)}\rightarrow \left\{\ba{lll} e^{\mathcal{R}_f^{(\a)}}\parallel g\parallel_{\infty}, & \b\rightarrow\infty, &\a~\mbox{ finite}\\
\left(\parallel f\parallel_{\infty}e^{\mathcal{R}_g^{(\b)}}\right)^{-1}, & \a\rightarrow\infty, &\b~\mbox{ finite}.\ea\right.
\eeq
For infinite discrete or continuous distribution functions $f$ and $g$, one can write
\beq\label{Cfg.xiib}
C^{(\a,\b)}_{(f,g)}\rightarrow \left\{\ba{lll} 0, & \b\rightarrow 0, &\a~\mbox{ finite}\\
	\infty, & \a\rightarrow 0, &\b~\mbox{ finite}.\ea\right.
\eeq  
If $\lim\limits_{\a\rightarrow 0}\mathcal{R}_f^{(\a)}=\ln \mathcal{O}(\Om_f^D)$, or $\ln\mathcal{L}(\Om_f^D)$, and $\lim\limits_{\b\rightarrow 0}\mathcal{R}_g^{(\b)}=\ln \mathcal{O}(\Om_g^D)$, or $\ln\mathcal{L}(\Om_g^D)$ exist, then one can improve the relation (\ref{Cfg.xiib}) as 
\beq
C^{(\a,\b)}_{(f,g)}\rightarrow \left\{\ba{lll} \ds\f{e^{\mathcal{R}_f^{(\a)}}}{ \mathcal{O}(\Om_g^D)}, & \b\rightarrow 0, &\a~\mbox{ finite}\\\\
\ds\f{ \mathcal{O}(\Om_f^D)}{e^{\mathcal{R}_g^{(\b)}}}, & \a\rightarrow 0, &\b~\mbox{ finite}.\ea\right.
\eeq
for discrete distributions, or 
\beq
C^{(\a,\b)}_{(f,g)}\rightarrow \left\{\ba{lll} \ds\f{e^{\mathcal{R}_f^{(\a)}}}{ \mathcal{L}(\Om_g^D)}, & \b\rightarrow 0, &\a~\mbox{ finite}\\
\ds\f{ \mathcal{L}(\Om_f^D)}{e^{\mathcal{R}_g^{(\b)}}}, & \a\rightarrow 0, &\b~\mbox{ finite}.\ea\right.
\eeq
for continuous distributions.

%	\noindent {\bf Upper bound by $L_p$ norm}
%\subsection{Scaling}
\noindent $\bullet$ (xi) $C^{(\a,\b)}_{(\bar{f},\bar{g})} =\left(\f{c}{a}\right)^DC^{(\a,\b)}_{(f,g)}$, for $\bar{f}({\bf r})=a^Df(a({\bf r-b}))$, $\bar{g}({\bf r})=c^Dg(c({\bf r-d}))$, where $D$ is the dimension of the system.\\
\noindent $\bullet$ (xii) $C^{(\a,\b)}_{(\bar{f},\bar{g})} =\left(\f{m}{n}\right)^{\f{D}{2}-1}C^{(\a,\b)}_{(f,g)}$, for $\bar{f}({\bf r})=\ds\sum_{i=0}^{n}f_i({\bf r})$, $\bar{g}({\bf r})=\ds\sum_{i=0}^{m}g_i({\bf r})$, where 
$f_i({\bf r})=n^{\f{D}{2}-1}f(\sqrt{n}({\bf r-a_i}))$, 
$g_i({\bf r})=m^{\f{D}{2}-1}f(\sqrt{m}({\bf r-b_i}))$, 
$\ds\int f_i{(\bf r)}d{\bf r}=\f{1}{n}, \int g_i({\bf r})d{\bf r}=\f{1}{m}$, 
$\ds\int \bar{f}({\bf r})d{\bf r}=\int\bar{g}({\bf r})d{\bf r}=\int f({\bf r})d{\bf r}=\int g({\bf r})d{\bf r}=1$, $D$ is the dimension of the system.

\subsection{Main theorems for near continuous property of RCR}
\begin{dfn}\label{def.delta.neigh}
		($\delta$-neighboring \cite{a.nagy.jmp,jc.angulo.jsd.epjd,nagy.j.stat.mec}). Let $f_1:M^D\rightarrow[0,1]$ and $f_2:M^D\rightarrow[0,1]$ be two density functions defined on a set $M^D$ in $D$-dimensional space and $\delta$ be a positive real number. Then the functions $f_1$ and $f_2$ are called the $\delta$-neighboring functions on $M^D$, if $\mathcal{L}\left(\left\{{\bf r}:|f_1({\bf r})-f_2({\bf r})|\ge \delta \right\}\right)=0$.
	\end{dfn}
\begin{dfn}\label{def.near.cont}
	(Near continuous \cite{a.nagy.jmp,jc.angulo.jsd.epjd,nagy.j.stat.mec}). A functional $T$, of density functions is said to be near continuous, if for every positive $\epsilon$ there exists a $\delta(\epsilon)>0$, such that $\mathcal{L}\left(\left\{{\bf r}:|f_1({\bf r})-f_2({\bf r})|\ge \delta \right\}\right)=0$ implies $|T(f_1)-T(f_2)|<\epsilon$.
\end{dfn}
Using definitions \ref{def.eff.r}, \ref{def.delta.neigh} and \ref{def.near.cont} we can define a theorem for near continuous property of R\'enyi complexity ratio for radial density functions.
\begin{theorem}\label{theorem.r} %3.1
	Let $(f_1,g_1)$ and $(f_2,g_2)$ be two pairs of radial density functions. If for a positive $\epsilon$, there exists a positive $\delta(\epsilon)$, such that $\mathcal{L}\left(S=\left\{r:\sqrt{\left(f_1(r)-f_2(r)\right)^2+\left(g_1(r)-g_2(r)\right)^2}\ge \delta \right\}\right)=0$, then  $|C^{(\a,\b)}_{(f_1,g_1)}-C^{(\a,\b)}_{(f_2,g_2)}|\rightarrow 0$ as $\delta\rightarrow 0$, for positive integers  $\a,\b$.
\end{theorem}
Proof. Let $F^{i\alpha}$ be the effective domain of the function $\left(f_i(r)\right)^{\a}$ with respect to $r$, for $i=1,2$. Since $f_1$ and $f_2$ are density functions then $0<\mathcal{L}(F^{i\alpha})<\infty$, for $i=1,2$, $\a\in\mathbb{N}$.

\noindent Now $\ds\int|f_1(r)-f_2(r)|r^2dr=\int_{S_1}|f_1(r)-f_2(r)|r^2dr+\int_{S'_1}|f_1(r)-f_2(r)|r^2dr$, where $S_1=\left(F^1\cup F^2\right)\cap S$, $S'_1=\left(F^1\cup F^2\right)- S$ and $F^i=F^{i1}$, $i=1,2$.\\ 
Then $\ds\int|f_1(r)-f_2(r)|r^2dr\le \delta\left[\sup S'_1\right]^2\mathcal{L}(S'_1)$ and  $\ds\int|f_1^{\alpha}(r)-f_2^{\alpha}(r)|r^2dr\le \delta\ds\sum\limits_{i=0}^{\alpha-1}\,I_{1}^{(i)}\,I_2^{(\alpha-1-i)}$, where $0\le I_j^{(i)}=\ds\int_{F^{j\alpha}-S}f_j^i(r)\,r^2dr<\infty $, for $i=0,1,\dots,\alpha$, $j=1,2$.\\
Therefore, $\ds\int \left(f_2^{\alpha}(r)-f_1^{\alpha}(r)\right)r^2dr\rightarrow 0$, as $\delta\rightarrow 0$, and\\ $\mathcal{R}_{f_2}^{(\alpha)}-\mathcal{R}_{f_1}^{(\alpha)}\le\f{1}{1-\alpha}\ln\left(\ds\f{\delta}{I_1^{(\a)}}\ds\sum\limits_{i=0}^{\alpha-1}\,I_{1}^{(i)}\,I_2^{(\alpha-1-i)}+1\right)\rightarrow 0,~\mbox{as~}\delta\rightarrow 0.
$\\
Hence $C^{(\a,\a)}_{(f_2,f_1)}=C^{\mathcal{R}_{f_2}^{(\alpha)}-\mathcal{R}_{f_1}^{(\alpha)}}\rightarrow 1$, as $\delta\rightarrow 0$. Similarly we can proof that $C^{(\b,\b)}_{(g_2,g_1)}=C^{\mathcal{R}_{g_1}^{(\b)}-\mathcal{R}_{g_2}^{(\b)}}\rightarrow 1$, as $\delta\rightarrow 0$.\\
Therefore, $\left|C^{(\a,\b)}_{(f_1,g_1)}-C^{(\a,\b)}_{(f_2,g_2)}\right|=C^{(\a,\b)}_{(f_1,g_1)}\left|1-\f{C^{(\a,\a)}_{(f_2,f_1)}}{C^{(\b,\b)}_{(g_2,g_1)}}\right|\rightarrow 0$, as $\delta\rightarrow 0$.\\
Using definitions \ref{def.eff.th}, \ref{def.delta.neigh} and \ref{def.near.cont} we can define another theorem.

\begin{theorem}\label{theorem.th}%3.2
	Let $(f_1,g_1)$ and $(f_2,g_2)$ be two pairs of density functions of polar angle $\theta$. If for a positive $\epsilon$, there exists a positive $\delta(\epsilon)$, such that $\mathcal{L}\left(S=\left\{\theta:\sqrt{\left(f_1(\theta)-f_2(\theta)\right)^2+\left(g_1(\theta)-g_2(\theta)\right)^2}\ge \delta \right\}\right)=0$, then $|C^{(\a,\b)}_{(f_1,g_1)}-C^{(\a,\b)}_{(f_2,g_2)}|\rightarrow 0$ as $\delta\rightarrow 0$, for positive integers  $\a,\b$.
\end{theorem}
Proof. Let $F^{i\alpha}$ be the effective domain of the function $\left(f_i(\theta)\right)^{\a}$ with respect to $\theta$, for $i=1,2$. Since $f_1$ and $f_2$ are density functions, then $0<\mathcal{L}(F^{i\alpha})<\infty$, for $i=1,2$, $\a\in\mathbb{N}$.\\

\noindent Now $\ds\int|f_1(\theta)-f_2(\theta)|\sin\theta d\theta=\int_{S_1}|f_1(\theta)-f_2(\theta)|\sin\theta d\theta+\int_{S'_1}|f_1(\theta)-f_2(\theta)|\sin\theta d\theta$, where $S_1=\left(F^1\cup F^2\right)\cap S$, $S'_1=\left(F^1\cup F^2\right)- S$ and $F^i=F^{i1}$, $i=1,2$.\\ 
Then $\ds\int|f_1(\theta)-f_2(\theta)|\sin\theta\,d\theta\le 2\delta$, and  $\ds\int|f_1^{\alpha}(\theta)-f_2^{\alpha}(\theta)|\sin\theta d\theta\le \delta\ds\sum\limits_{i=0}^{\alpha-1}\,I_{1}^{(i)}\,I_2^{(\alpha-1-i)}$,\\ 
where $0\le I_j^{(i)}=\ds\int_{F^{j\alpha}-S}f_j^i(\theta)\,\sin\theta d\theta<\infty$, for $i=0,1,\dots,\alpha$, $j=1,2$.\\
Therefore, $\delta\rightarrow 0$ implies $\ds\int \left(f_2^{\alpha}(\theta)-f_1^{\alpha}(\theta)\right)\sin\theta\,d\theta\rightarrow 0$, \\
and $\mathcal{R}_{f_2}^{(\alpha)}-\mathcal{R}_{f_1}^{(\alpha)}\le\f{1}{1-\alpha}\ln\left(\ds\f{\delta}{I_1^{(\a)}}\ds\sum\limits_{i=0}^{\alpha-1}\,I_{1}^{(i)}\,I_2^{(\alpha-1-i)}+1\right)\rightarrow 0$, as $\delta\rightarrow 0$.\\
Hence $C^{(\a,\a)}_{(f_2,f_1)}=C^{\mathcal{R}_{f_2}^{(\alpha)}-\mathcal{R}_{f_1}^{(\alpha)}}\rightarrow 1$, as $\delta\rightarrow 0$. Similarly we can proof that $C^{(\b,\b)}_{(g_2,g_1)}=C^{\mathcal{R}_{g_2}^{(\b)}-\mathcal{R}_{g_1}^{(\b)}}\rightarrow 1$, as $\delta\rightarrow 0$.\\
Therefore, $\left|C^{(\a,\b)}_{(f_1,g_1)}-C^{(\a,\b)}_{(f_2,g_2)}\right|=C^{(\a,\b)}_{(f_1,g_1)}\left|1-\f{C^{(\a,\a)}_{(f_2,f_1)}}{C^{(\b,\b)}_{(g_2,g_1)}}\right|\rightarrow 0$, as $\delta\rightarrow 0$.\\
Similarly using the definitions \ref{def.eff.ph}, \ref{def.delta.neigh} and \ref{def.near.cont}, we can define a theorem for near continuous property of R\'enyi complexity ratio of density functions of $\phi$.

\begin{theorem}\label{theorem.ph}%3.3
	Let $(f_1,g_1)$ and $(f_2,g_2)$ be two pairs of density functions azimuthal angle $\phi$ defined on $(0,2\pi)$. If for a positive $\epsilon$, there exists a positive $\delta(\epsilon)$, such that $\mathcal{L}\left(S=\left\{\phi:\sqrt{\left(f_1(\phi)-f_2(\phi)\right)^2+\left(g_1(\phi)-g_2(\phi)\right)^2}\ge \delta \right\}\right)=0$, then  $|C^{(\a,\b)}_{(f_1,g_1)}-C^{(\a,\b)}_{(f_2,g_2)}|\rightarrow 0$ as $\delta\rightarrow 0$, for positive integers  $\a,\b$.
\end{theorem}
Proof. Let $F^{i\alpha}$ be the effective domain of the function $\left(f_i(\phi)\right)^{\a}$ with respect to $\phi$, for $i=1,2$. Since $f_1$ and $f_2$ are density functions then $0<\mathcal{L}(F^{i\alpha})<\infty$, for $i=1,2$, $\a\in\mathbb{N}$.\\

\noindent Now $\ds\int|f_1(\phi)-f_2(\phi)|d\phi=\int_{S_1}|f_1(\phi)-f_2(\phi)|d\phi+\int_{S'_1}|f_1(\phi)-f_2(\phi)|d\phi$, where $S_1=\left(F^1\cup F^2\right)\cap S$, $S'_1=\left(F^1\cup F^2\right)- S$ and $F^i=F^{i1}$, $i=1,2$.\\ 
Then $\ds\int|f_1(\phi)-f_2(\phi)|\,d\phi\le 2\pi\delta$ and  $\ds\int|f_1^{\alpha}(\phi)-f_2^{\alpha}(\phi)|\,d\phi\le \delta\ds\sum\limits_{i=0}^{\alpha-1}\,I_{1}^{(i)}\,I_2^{(\alpha-1-i)}$, where $0\le I_j^{(i)}=\ds\int_{F^{j\alpha}-S}f_j^i(\phi)\,d\phi<\infty$, for $i=0,1,\dots,\alpha$, $j=1,2$.\\
Therefore $\delta\rightarrow 0$ implies $\ds\int \left(f_2^{\alpha}(\phi)-f_1^{\alpha}(\phi)\right)\,d\phi\rightarrow 0$.\\ 
Now $\mathcal{R}_{f_2}^{(\alpha)}-\mathcal{R}_{f_1}^{(\alpha)}\le\f{1}{1-\alpha}\ln\left(\ds\f{\delta}{I_1^{(\a)}}\ds\sum\limits_{i=0}^{\alpha-1}\,I_{1}^{(i)}\,I_2^{(\alpha-1-i)}+1\right)\rightarrow 0$, as $\delta\rightarrow 0$.\\
Hence $C^{(\a,\a)}_{(f_2,f_1)}=C^{\mathcal{R}_{f_2}^{(\alpha)}-\mathcal{R}_{f_1}^{(\alpha)}}\rightarrow 1$, as $\delta\rightarrow 0$. Similarly one can proof that $C^{(\b,\b)}_{(g_2,g_1)}=C^{\mathcal{R}_{g_2}^{(\b)}-\mathcal{R}_{g_1}^{(\b)}}\rightarrow 1$ as $\delta\rightarrow 0$.\\
Therefore, $\left|C^{(\a,\b)}_{(f_1,g_1)}-C^{(\a,\b)}_{(f_2,g_2)}\right|=C^{(\a,\b)}_{(f_1,g_1)}\left|1-\f{C^{(\a,\a)}_{(f_2,f_1)}}{C^{(\b,\b)}_{(g_2,g_1)}}\right|\rightarrow 0$, as $\delta\rightarrow 0$.\\
Moreover, using the definitions \ref{def.eff.vecr}, \ref{def.delta.neigh} and \ref{def.near.cont} we can define a theorem for near continuous property of the R\'enyi complexity ratio of density functions in spherical coordinates system.

\begin{theorem}\label{theorem.vecr}%3.4
	Let $(f_1,g_1)$ and $(f_2,g_2)$ be two pairs of joint density functions of ${\bf r}$, defined on $\mathbb{R}^g=(0,\infty)\times(0,\pi)\times(0,2\pi)$. If for a positive $\epsilon$, there exists a positive $\delta(\epsilon)$, such that\\ $\mathcal{L}\left(S=\left\{{\bf r}:\sqrt{\left(f_1({\bf r})-f_2({\bf r})\right)^2+\left(g_1({\bf r})-g_2({\bf r})\right)^2}\ge \delta \right\}\right)=0$, then  $|C^{(\a,\b)}_{(f_1,g_1)}-C^{(\a,\b)}_{(f_2,g_2)}|\rightarrow 0$ as $\delta\rightarrow 0$, for positive integers  $\a,\b$.
\end{theorem}
Proof. Let $f_i({\bf r})=f_{ir}(r)f_{i\theta}(\theta)f_{i\phi}(\phi)$, be defined in $\mathbb{R}^g$. Moreover $F_{r}^{i\alpha}$, $F_{\theta}^{i\alpha}$ and $F_{\phi}^{i\alpha}$ be the effective domains of $f_{ir}$, $f_{i\theta}$ and $f_{i\phi}$ with respect to $r$, $\theta$ and $\phi$ respectively, for $i=1,2$. Since $f_1$ and $f_2$ are density functions then $0<\mathcal{L}(F_{r}^{i\alpha}),\mathcal{L}(F_{\theta}^{i\alpha}),\mathcal{L}(F_{\phi}^{i\alpha})<\infty$, for $i=1,2$, $\a\in\mathbb{N}$. Then $\mathcal{F}_j=F_r^{j}\times F_{\theta}^j\times F_{\phi}^j$ is the effective domain of $f_j$, for $F_r^j=F_r^{j1},F_{\theta}^j=F_{\theta}^{j1},F_{\phi}^j=F_{\phi}^{j1}$, for $j=1,2$.\\

\noindent Now $\ds\int|f_1({\bf r})-f_2({\bf r})|d{\bf r}=\int_{S_1}|f_1({\bf r})-f_2({\bf r})|d{\bf r}+\int_{S'_1}|f_1({\bf r})-f_2({\bf r})|d{\bf r}$, where $S_1=\mathcal{F}_1\cup \mathcal{F}_2\cap S$, $S'_1=\mathcal{F}_1\cup\mathcal{F}_2- S$.\\ 
Then $\ds\int|f_1({\bf r})-f_2({\bf r})|\,d{\bf r}\le 4\pi\delta\left[\sup F_r^1\cup F_r^2\right]^2\mathcal{L}(F_r^1\cup F_r^2)$ and  $\ds\int|f_1^{\alpha}({\bf r})-f_2^{\alpha}({\bf r})|\, d{\bf r}\le \delta\ds\sum\limits_{i=0}^{\alpha-1}\,I_{1}^{(i)}\,I_2^{(\alpha-1-i)}$, where $0\le I_j^{(i)}=\ds\int_{\mathcal{F}_j-S}f_j^i({\bf r})\, d{\bf r}<\infty$, for $i=0,1,\dots,\alpha-1$, $j=1,2$.\\

\noindent Therefore, $\delta\rightarrow 0$ implies $\ds\int \left(f_2^{\alpha}({\bf r})-f_1^{\alpha}({\bf r})\right)\,d{\bf r}\rightarrow 0$,\\ 
and $\mathcal{R}_{f_2}^{(\alpha)}-\mathcal{R}_{f_1}^{(\alpha)}\le\f{1}{1-\alpha}\ln\left(\f{\delta\ds\sum\limits_{i=0}^{\alpha-1}\,I_{1}^{(i)}\,I_2^{(\alpha-1-i)}}{\ds\int f_1^{\a}({\bf r})\,d{\bf r}}+1\right)\rightarrow 0$, as $\delta\rightarrow 0$.\\
Thus we can write,  $C^{(\a,\a)}_{(f_2,f_1)}=C^{\mathcal{R}_{f_2}^{(\alpha)}-\mathcal{R}_{f_1}^{(\alpha)}}\rightarrow 1$, as $\delta\rightarrow 0$. Similarly we can proof that $C^{(\b,\b)}_{(g_2,g_1)}=C^{\mathcal{R}_{g_1}^{(\b)}-\mathcal{R}_{g_2}^{(\b)}}\rightarrow 1$, as $\delta\rightarrow 0$.\\
Therefore, $\left|C^{(\a,\b)}_{(f_1,g_1)}-C^{(\a,\b)}_{(f_2,g_2)}\right|=C^{(\a,\b)}_{(f_1,g_1)}\left|1-\f{C^{(\a,\a)}_{(f_2,f_1)}}{C^{(\b,\b)}_{(g_2,g_1)}}\right|\rightarrow 0$, as $\delta\rightarrow 0$.

For a $D$-dimensional central potential total density function $\rho({\bf r})_{n,\ell,\{\mu\}}$ can be written as a product of hyper-radial density $(d_{n,\ell}(r)=|R_{n,\ell}|^2)$ and hyper-spherical density function $(|Y_{\ell,\{\mu\}}|^2)$, where 
	\beq\label{Ylm}
	Y_{\ell,\{\mu\}}=N_{\ell,\{\mu\}}e^{im\theta_{D-1}}\prod\limits_{j=1}^{D-2}C_{\mu_j-\mu_{j+1}}^{\a_j+\mu_j+1}(\cos\theta_j)(\sin\theta_j)^{\mu_j+1},
	\eeq
	and $R_{n,\ell}$ corresponds to wave function of hyper-radial Schr\"odinger equation, $C_{i}^{j}$ is the Gegenbauer polynomial of degree $i$ with parameter $j$, $N_{\ell,\{\mu\}}$ provides the normalization constant and $(\ell,\{\mu\})=(\mu_1,\mu_2,...,\mu_{D-1})$, $\ell=\mu_1\ge\mu_2\ge...\ge\mu_{D-2}\ge|\mu_{D-1}|=|m|$, $\ell=0,1,2,...$, $m=0,\pm1,\pm2,...$, $\a_j=(D-j-1)/2$. In this case the measure $d{\bf r}=d^Dr=r^{D-1}dr\,d\Om_{D-1}$, where $d\Om_{D-1}=\left[\prod\limits_{j=1}^{D-2}\left(\sin\theta_j\right)^{2\a_j}\,d\theta_j\right]d\theta_{D-1}$. For $D=3$, the hyperspherical density functions $|Y_{\ell,{\mu}}|^2$ has a simple form which is defined in Eq. (\ref{Ylm3}) and $d{\bf r}=r^2\sin\theta\,dr\,d\theta\,d\phi$. Therefore, for central potential total density function is separable and for non-central potential it may not be separable. If a total density function $\rho({\bf r})$ in a $D$-dimensional quantum system has effective domain $\Om$, then $\ds\int_{\Om}\rho({\bf r})d{\bf r}=1$, $\ds\int_{\Om}\left[\rho({\bf r})\right]^{\a}d{\bf r}<\infty$ for $\a=0,1,2,....$. Therefore, in a similar manner one can extend theorem \ref{theorem.vecr} for a $D$-dimensional quantum system in more general circumstances.

\begin{theorem}\label{theorem.muld}%3.5
	Let $(f_1,g_1)$ and $(f_2,g_2)$ be two pairs of joint density functions defined on $M^D$ have effective domains. If for a positive $\epsilon$, there exists a positive $\delta(\epsilon)$, such that $\mathcal{L}\left(\left\{{\bf r}:\sqrt{\left(f_1({\bf r})-f_2({\bf r})\right)^2+\left(g_1({\bf r})-g_2({\bf r})\right)^2}\ge \delta \right\}\right)=0$, then  $|C^{(\a,\b)}_{(f_1,g_1)}-C^{(\a,\b)}_{(f_2,g_2)}|\rightarrow 0$ as $\delta\rightarrow 0$, for positive integers  $\a,\b$.
\end{theorem}
Proof. Let $\mathcal{F}_j$ be the $D$-dimensional effective domain of $f_j({\bf r})$ with respect to ${\bf r}$ for $j=1,2$ defined in $M^D$. Since $f_1$, $f_2$ are density functions, we have $0<\ds\int_{\mathcal{F}_1\cup\mathcal{F}_2-S}d{\bf r}<\infty$, and $0\le I_j^{(i)}=\ds\int_{\mathcal{F}_j-S}\left(f_j({\bf r})\right)^i\, d{\bf r}<\infty$, for $i=0,1,\dots,\alpha-1$, $j=1,2$.

Then $\ds\int|f_1({\bf r})-f_2({\bf r})|\,d{\bf r}\le \delta\mathcal{V}$ and  $\ds\int|f_1^{\alpha}({\bf r})-f_2^{\alpha}({\bf r})|\, d{\bf r}\le \delta\ds\sum\limits_{i=0}^{\alpha-1}\,I_{1}^{(i)}\,I_2^{(\alpha-1-i)}$, where $\mathcal{V}$ is the volume of the effective domain.\\

\noindent Therefore $\delta\rightarrow 0$ implies $\ds\int \left(f_2^{\alpha}({\bf r})-f_1^{\alpha}({\bf r})\right)\,d{\bf r}\rightarrow 0$,\\ 
and $\mathcal{R}_{f_2}^{(\alpha)}-\mathcal{R}_{f_1}^{(\alpha)}\le\f{1}{1-\alpha}\ln\left(\f{\delta\ds\sum\limits_{i=0}^{\alpha-1}\,I_{1}^{(i)}\,I_2^{(\alpha-1-i)}}{\ds\int f_1^{\a}({\bf r})\,d{\bf r}}+1\right)\rightarrow 0$, as $\delta\rightarrow 0$.\\
Thus we can write,  $C^{(\a,\a)}_{(f_2,f_1)}=C^{\mathcal{R}_{f_2}^{(\alpha)}-\mathcal{R}_{f_1}^{(\alpha)}}\rightarrow 1$, as $\delta\rightarrow 0$. Similarly, we can write that $C^{(\b,\b)}_{(g_2,g_1)}=C^{\mathcal{R}_{g_1}^{(\b)}-\mathcal{R}_{g_2}^{(\b)}}\rightarrow 1$, as $\delta\rightarrow 0$.\\
Therefore, $\left|C^{(\a,\b)}_{(f_1,g_1)}-C^{(\a,\b)}_{(f_2,g_2)}\right|=C^{(\a,\b)}_{(f_1,g_1)}\left|1-\f{C^{(\a,\a)}_{(f_2,f_1)}}{C^{(\b,\b)}_{(g_2,g_1)}}\right|\rightarrow 0$, as $\delta\rightarrow 0$.\\

If $f_1$ and $f_2$ are $\d$-neighboring total density functions, then their corresponding reduced density functions are also $\d$-neighboring. Similarly, if $(f_1,g_1)$ and $(f_2,g_2)$ are two pairs of total density functions satisfy near continuous property of RCR, then their corresponding reduced density functions will be satisfy the near continuous property of RCR. Therefore, if $(f_1,g_1)$ and $(f_2,g_2)$ are two pairs of density functions defined on a closed and bounded domain (compact set) satisfy $\d$-neighboring property then, they satisfy the near continuous property of RCR.
\subsection{Example of near continuous property of RCR}
Now one can say that, the R\'enyi complexity ratio satisfies the near continuous property, for $\a,\b\in\mathbb{N}$ with help of the definitions \ref{def.eff.r}, \ref{def.eff.th}, \ref{def.eff.ph}, \ref{def.eff.vecr}, \ref{def.delta.neigh}, \ref{def.near.cont} and the theorems \ref{theorem.r}, \ref{theorem.th}, \ref{theorem.ph}, \ref{theorem.vecr} and \ref{theorem.muld}. It is difficult to proof above theorems for non integral values of $\a$ and $\b$. One can proves and verifies near continuous property of RCR for positive real values of $\a$ and $\b$ by counter examples of density functions, such as step function and uniform density functions \cite{a.nagy.jmp,a.nagy.chaos,jc.angulo.jsd.epjd,jsd.pla2016,nagy.j.stat.mec} and so on.\\ 
Let us consider two pairs of $(f_1,g_1)$ and $(f_2,g_2)$ density functions defined on a $D$-dimensional space by
\beq
\ba{l}
f_1({\bf r})=\ds\left\{\ba{ll}\f{1-\delta_1}{C_D},&|{\bf r}|<1\\
\ds\f{\delta_1}{C_D(B^D-1)},&1<|{\bf r}|<B\\
0,&\mbox{elsewhere}, \ea\right.
\ea
\eeq
\beq
\ba{l}
g_1({\bf r})=\ds\left\{\ba{ll}\f{1-\delta'_1}{C_D},&|{\bf r}|<1\\
\ds\f{\delta'_1}{C_D(B^D-1)},&1<|{\bf r}|<B\\
0,&\mbox{elsewhere}, \ea\right.
\ea
\eeq
and
\beq
\ba{l}
f_2({\bf r})=g_2({\bf r})=\left\{\ba{ll}\f{1}{C_D},&|{\bf r}|<1\\
0,&\mbox{elsewhere}, \ea\right.
\ea
\eeq
where $C_D=\f{2\pi^{\f{D}{2}}}{D\G(\f{D}{2})}$, $B>1$, $0<\delta_1,\delta'_1<1$. Then $(f_1,f_2)$ and $(g_1,g_2)$ are pairs of neighboring functions. 
Therefore, one can obtain 
\beq
\mathcal{R}_{f_1}^{(\a)}=\ds\f{1}{1-\a}\ln\left[(1-\delta_1)^{\a}+\f{\delta_1^{\a}}{(B^D-1)^{\a-1}}\right]+\ln C_D,
\eeq 
\beq
\mathcal{R}_{g_1}^{(\a)}=\ds\f{1}{1-\a}\ln\left[(1-\delta'_1)^{\a}+\f{\delta_1^{'\a}}{(B^D-1)^{\a-1}}\right]+\ln C_D,
\eeq 
and 
\beq
\mathcal{R}_{f_2}^{(\a)}=\mathcal{R}_{g_2}^{(\b)}=\ln C_D.
\eeq 
Hence, it follows that
\beq
\lim\limits_{\delta_1\rightarrow 0} \mathcal{R}_{f_1}^{(\a)}=\mathcal{R}_{f_2}^{(\a)}=\ln C_D,
\eeq
and
\beq
\lim\limits_{\delta'_1\rightarrow 0} \mathcal{R}_{g_1}^{(\a)}=\mathcal{R}_{g_2}^{(\a)}=\ln C_D.
\eeq 
Therefore, 
\beq
\lim\limits_{\delta_1\rightarrow 0}\lim\limits_{\delta'_1\rightarrow 0} C_{(f_1,g_1)}^{(\a,\b)}=\lim\limits_{\delta'_1\rightarrow 0}\lim\limits_{\delta_1\rightarrow 0} C_{(f_1,g_1)}^{(\a,\b)}=C_{(f_2,g_2)}^{(\a,\b)}=1.
\eeq
\subsection{Extremal property of RCR}
Let $f$ and $g$ be two density functions define on $\Om^D$, such that
\beq
\ba{l}
f({\bf r})=\ds\sum\limits_i\,p_i\,\chi_{\Om_i},~g({\bf r})=\ds\sum\limits_i\,p_i\,\chi_{\Om_i},\\
\ds\sum\limits_i\,p_i\,m_i=1,\ds\sum\limits_i\,q_i\,m_i=1,\\
\ea
\eeq
where $m_i=\mathcal{L}(\Om_i)$, $i=1,2,...,n$. Then
\beq
\ba{l}
\ds\int_{\Om^D}f^{\a}d{\bf r}=\ds\sum\limits_i\,p_i^{\a}\,m_i,\ds\int_{\Om^D}g^{\b}d{\bf r}=\ds\sum\limits_i\,q_i^{\b}\,m_i.
\ea 
\eeq 
Let us assume that $\widetilde{C}_{(f,g)}^{(\a,\b)}=\ln C_{(f,g)}^{(\a,\b)}$. Then one can write 
\beq
\widetilde{C}=\f{1}{1-\a}\ln\left(\ds\sum\limits_i\,p_i^{\a}\,m_i\right)-\f{1}{1-\b}\ln\left(\ds\sum\limits_i\,q_i^{\b}\,m_i\right).
\eeq 
Then from the variation of $\widetilde{C}$ with respect to $m_i$ 
\beq\label{Eq.wrt.mi}
\f{p_i^{\a}}{q_i^{\b}}=\f{(1-\a)\ds\sum\limits_i\,p_i^{\a}\,m_i}{(1-\b)\ds\sum\limits_i\,q_i^{\b}\,m_i}.
\eeq
Similarly, from the variation of $\widetilde{C}$ with respect to $p_i$ and $q_i$, one can obtain $\a=0$, $g$ uniform and $\b=0$, $f$ uniform respectively. For non zero values of $\a$ and $\b$, the second order variation of $\widetilde{C}$ with respect to $p_i$ and $q_i$ are considered. Thus from the second order variation of $\widetilde{C}$, one obtains
\beq\label{Eq.wrt.pi.qi}
\ba{l}
p_i^{\a}m_i=\f{\a-1}{\a}\ds\sum\limits_i\,p_i^{\a}m_i,\\
q_i^{\b}m_i=\f{\b-1}{\b}\ds\sum\limits_i\,q_i^{\b}m_i.
\ea
\eeq 
From Eqs.(\ref{Eq.wrt.mi}) and (\ref{Eq.wrt.pi.qi}), it is to be noted that $\a,\b>1$. \\
Therefore,\\ 
(i) uniform $f$, arbitrary density $g$, with $\a>0$ \& $\b=0$,\\ 
(ii) uniform $g$, arbitrary density $f$, with $\a=0$ \& $\b>0$, and\\ 
(iii) uniform $f$, $g$ with $\a=\b =\ds\f{\mathcal{L}(\Om^D)}{\mathcal{L}(\Om^D-\Om_i)}$, $i=1,2,...,n$\\ 
are solutions of (\ref{Eq.wrt.mi}) and (\ref{Eq.wrt.pi.qi}). Now if, $f$ and $g$ be two uniform distributions, then $C_{(f,g)}^{(\a,\b)}$ does not depend on $\a$ and $\b$, but it violet the extremality conditions. Finally $C_{(f,g)}^{(\a,\b)}$ has extremal values.
\subsection{Properties of RCR for isospectral quantum systems}
If $(\psi,E)$ and $(\widehat{\psi},\widehat{E})$ are two pairs of eigen functions and eigen values of the Hamiltonians $H_A=A^{\dag}A$ and $H_B=B^{\dag}B$ respectively, where $A(B)$ and $A^{\dag}(B^{\dag})$ are the annihilation and creation operators, $AA^{\dag}=BB^{\dag}$ but $A^{\dag}A\ne B^{\dag}B$. Then for two isospectral \cite{SUSY} density functions $\rho=|\psi|^2$ of $H_A$ and $\widehat{\rho}(\lam)=|\widehat{\psi}(\lam)|^2$ of $H_B$ one can write\\
(i) $\ds C^{(\a,\b)}_{(\widehat{\rho},\widehat{\rho})}(\lam),C^{(\a,\b)}_{(\widehat{\rho},\rho)}(\lam), C^{(\a,\b)}_{(\rho,\widehat{\rho})}(\lam)\rightarrow C^{(\a,\b)}_{(\rho,\rho)}$ as $|\lam|\rightarrow\infty$, and\\
(ii) $C^{(\a,\a)}_{(\widehat{\rho},\rho)}(\lam),C^{(\a,\a)}_{(\rho,\widehat{\rho})}(\lam)\rightarrow 1$ as $|\lam|\rightarrow\infty$.\\ 
where $\lam$ is the isospectral parameter. In this paper, $\ds C^{(\a,\b)}_{(\widehat{\rho},\widehat{\rho})}(\lam),C^{(\a,\b)}_{(\widehat{\rho},\rho)}(\lam), C^{(\a,\b)}_{(\rho,\widehat{\rho})}(\lam)$  \& $C^{(\a,\b)}_{(\rho,\rho)}$ will be found and discussed about these two properties. 

\subsection{Generalized R\'enyi complexity and shape R\'enyi complexity}\label{SR.GR}
The R\'enyi complexity ratio $C^{(\a,\b)}_{(\rho,\rho)}$ is called generalized R\'enyi complexity of order $(\a,\b)$. It is denoted by $C^{(\a,\b)}$ and defined by \cite{a.nagy.jmp,a.nagy.chaos,jc.angulo.jsd.epjd,jsd.pla2016,DN.ijmpa}
\beq
C^{(\a,\b)}=C^{(\a,\b)}_{(\rho,\rho)}=e^{\mathcal{R}_{\rho}^{(\a)}-\mathcal{R}_{\rho}^{(\b)}},~\a,\b>0.
\eeq
The shape R\'enyi complexity \cite{jc.anguloCPL,a.nagy.IRP} of order $\a$ is defined by 
\beq
C^{(\a)}=C^{(\a,2)}_{(\rho,\rho)}.
\eeq 
In the limiting case $\left(\a\rightarrow 1\right)$, R\'enyi entropy $\mathcal{R}_{\rho}^{(\a)}$ reduces to Shannon entropy, therefore, the modified or shape LMC complexity \cite{lmc.continuous,Yamano.JMP2004,lmc.biophys} is defined by $C^{(1,2)}_{(\rho,\rho)}$. Hence, one can write
\beq
\ba{l}\label{GR.Inequality}
C^{(\a,\b)}_{(f,g)}=\left\{\ba{ll}GRC, &\mbox{~if~}f=g,\\ 
SRC, &\mbox{~if~}f=g,\b=2,\\
C^{LMC},  &\mbox{~if~}f=g,\b=2,\a=1.
\ea \right\}.
\ea 
\eeq
A special case is that, $\ds\ln \left[C_{(f,f)}^{(1,2)}\right]$ represents the structural entropy of $f$. 
\section{Application}\label{section.appl}
%{\bf Central potential}\\
In this section R\'enyi entropy, R\'enyi complexity ratio, generalized R\'enyi complexity and shape R\'enyi complexity will be discussed. To do so pseudoharmonic oscillator and a family of isospectral potentials have been considered. Let us consider the pseudoharmonic oscillator potential \cite{pshopEx1.2,pshop.E.Kasap,Flugge,pshopEx2,pshop2.R.Sever.theo.chem,pshop3.R.Sever.jmc2007,pshop4,pshop5.R.Sever.jmc2012,pshop6.kds.oyewumi.jmc2012,pshop7.R.Sever.jmc2012.1973,pshop8.H.Hassanbadi,pshop9,pshop10.F.Chand.Pramana,DN.ijqc.pho} of the form
\beq\label{pseudo.3D}
V^{3D}(r)=D_e\left(\f{r}{r_e}-\f{r_e}{r}\right)^2,
\eeq
where $D_e=D_0+\ds\f{\hbar\om_e}{2}$, $D_0$ is the chemical dissociation energy, $\om_e$ is called harmonic vibrational parameters, $r$ is the internuclear distance between diatomic molecules. The pseudoharmonic oscillator potential is solvable for any angular momentum number $\ell$. It is minimum at point $r=r_e$ and it behaves like harmonic oscillator. It is one of the most important molecular potential for diatomic molecules \cite{pho.import}. It is used to describe interaction of some diatomic molecules \cite{pshop4,pshop6.kds.oyewumi.jmc2012}. 

%{\bf Under which circumstances the pseudoharmonic potential and the isospectral family approximately describe a diatomic molecule?}
\subsection{Solutions of pseudoharmonic oscillator and a family of isospectral potentials. }
 Then a family of isospectral potentials of the pseudoharmonic oscillator in spherical coordinates is \cite{DN.ijqc.qsi}
\beq\label{pseudo.iso.3D}
\widehat{V}^{3D}(r)=D_e\left(\f{r}{r_e}-\f{r_e}{r}\right)^2-\f{\hbar^2}{\mu}\f{d^2}{dr^2}\left[\ln\left(\lam+\mathcal{I}\right)\right],
\eeq
where
\beq
\ba{ll}
\mathcal{I}&=\ds\f{z^{L+\f{3}{2}}}{\G(L+\f{3}{2})}\ds\sum_{j=0}^{\infty}\f{(-z)^j}{j!(L+\f{3}{2}+j)},~z=ar^2,\\
L&=\ds-\f{1}{2}+\sqrt{\ell(\ell+1)+\f{1}{4}+a^2r_e^4},\\
a&=\f{\sqrt{2\mu D_e}}{\hbar r_e},
\ea 
\eeq
$\ell$ is the angular momentum number. Therefore, the wave solution and the ro-vibrational energy of the Schr\"odinger equation for pseudoharmonic oscillator potential (\ref{pseudo.3D}) are respectively \cite{pshop.E.Kasap,DN.ijqc.pho} 
\beq\label{sol.pseudo}
\ba{r}
\psi_{n,\ell,m}({\bf r})=\sqrt{a}N_{n}e^{-\f{1}{2}ar^2}\,(\sqrt{a}r)^{L}\ds L_{n}^{L+\f{1}{2}}\left(ar^2\right)Y_{\ell,m}(\theta,\phi),
\ea
\eeq
and
\beq
E_{n,\ell,m}^{3D}=\hbar\om_r(4n+2L+3)-2D_e,
\eeq 
where $L_{n}^{L+\f{1}{2}}\left(ar^2\right)$ is the associate Laguerre polynomial \cite{Ryzhik} of degree $n$ in $ar^2$ with parameter $L+\f{1}{2}$,
\beq
N_{n}=\sqrt{\f{n!\,2\sqrt{a}}{\G(n+L+\f{3}{2})}}
\eeq
is the normalization constants and $\om_r=\sqrt{\f{D_e}{2\mu r_e^2}}$. The harmonic spherical function is defined by \cite{Flugge}
\beq\label{Ylm3}
Y_{\ell,m}(\theta,\phi)=\left[\f{(2\ell+1)(\ell-|m|)!}{4\pi(\ell+|m|)!}\right]^{\f{1}{2}}P_{\ell}^{|m|}(\cos\theta)\,e^{im\phi},
\eeq
where $P_{\ell}^{|m|}(\cos\theta)$ is the associate Legendre polynomial \cite{Ryzhik} of degree $\ell$ in $\cos\theta$ and parameter $m$. 
On the other hand the wave solution and the ro-vibrational energy of a family of isospectral potentials (\ref{pseudo.iso.3D}) are respectively \cite{DN.IJQC.Renyi,DN.jmc,DN.ijqc.iso,DN.ijqc.qsi}
\beq\label{sol.pseudo.iso}
\ba{r}
\widehat{\psi}_{n,\ell,m}({\bf r},\lam)=\widehat{C}_n\sqrt{\Om(r,\lam)}\Phi_{n}(r)\,Y_{\ell,m}(\theta,\phi),%\\
%&=\widehat{C}_n\ds\f{\sqrt{2a^{\f{3}{2}}\G(L+\f{3}{2})}\,(\sqrt{a}r)^{L}\,e^{-\f{1}{2}ar^2}}{(\lam+1)\G(L+\f{3}{2})-\G(L+\f{3}{2},ar^2)}\,\Phi_{n}(r)\,Y_{l,m}(\theta,\phi),
\ea
\eeq
and
\beq
\widehat{E}^{3D}_{n,\ell,m}=\hbar\om_r(4n+2L+3)-2D_e,
\eeq  
where 
\begin{equation}
\ba{ll}
\Om(r,\lam)&=\ds\f{2a^{\f{3}{2}}\G(L+\f{3}{2})\,(ar^2)^{L}\,e^{-ar^2}}{\left((\lam+1)\G(L+\f{3}{2})-\G(L+\f{3}{2},ar^2)\right)^2},
\ea
\end{equation}
and
\begin{equation}
\ba{ll}
\Phi_{n}(r)&=1,~n=0,\\
&=\displaystyle\left[\lam + 1-\f{\G\left(L+\f{3}{2},ar^2\right)}{\G\left(L+\f{3}{2}\right)}\right]L_{n}^{L+\f{1}{2}}\left(ar^2\right)\\
&- \frac{(\sqrt{a}r)^{2L+3}e^{-ar^2}}{n\G\left(L+\f{3}{2}\right)}L_{n-1}^{L+\f{3}{2}}\left(ar^2\right),~~n=1,2,\cdots,
\ea
\end{equation}
are orthogonal functions with normalization constants 
\beq
{\widehat C}_0=\displaystyle\sqrt{\lam(\lam+1)},~~{\widehat C}_n=\sqrt{\f{n!\,\G(L+\f{3}{2})}{\G(n+L+\f{3}{2})}},n=1,2,\dots.
\eeq 
 
\subsection{R\'enyi entropy}
The nature of the density functions are important for finding the information theoretic measures. Now the normalized density functions of the states (\ref{sol.pseudo}) and (\ref{sol.pseudo.iso}) are defined by 
\beq\label{rho.n}
\ba{r}
\rho_{n,\ell,m}({\bf r})=aN_{n}^2\,e^{-ar^2}(ar^2)^L\left[L_{n}^{L+\f{1}{2}}\left(ar^2\right)\right]^2|Y_{\ell,m}(\theta,\phi)|^2,
\ea
\eeq 
and
\beq\label{rho.iso.n}
\ba{r}
\widehat{\rho}_{n,\ell,m}({\bf r},\lam)=\f{2a^{\f{3}{2}}\G(L+\f{3}{2})\widehat{C}^2_n\,(ar^2)^{L}\,e^{-ar^2}\Phi^2_{n}(r)|Y_{\ell,m}(\theta,\phi)|^2}{\left\{(\lam+1)\G(L+\f{3}{2})-\G(L+\f{3}{2},ar^2)\right\}^2}.
\ea
\eeq 
Therefore, the R\'enyi entropy of (\ref{rho.n}) of positive integral order $\a$ is defined by 
\beq
\ba{l}
%	\begin{eqnarray}
\mathcal{R}_{n,\ell,m}^{(\a)}=\f{1}{1-\a}\ln\left[\ds C\,A_0\left(\bar{\mu}_1,0,2\a,\{n\},\left\{L+\f{1}{2}\right\},\left\{\f{1}{\a}\right\}\right)J_{2,(\ell,m)}^{(\a)}\right],~n=0,1,2,\dots,
\ea
\eeq
where
%\begin{widetext}
\beqas
\ba{ll}
A_p\left(\mu,\b,2\a,\{m_i\},\left\{a_i\right\},\left\{t_i\right\}\right)&=\tiny{(\b+1)_{\mu}\binom{m_1+a_1}{m_1}\dots\binom{m_{2\a}+a_{2\a}}{a_{2\a}} F_A^{(2\a+1)}\left(\ba{r}\mu+\b+1;~\overbrace{-m_1,...,-m_{2\a}}^{2\a},~-p\\~~~\underbrace{a_1+1,...,a_{2\a}+1}_{2\a},~\beta+1\ea;~\overbrace{t_1,...,t_{2\a}}^{2\a},~1\right)},
\ea
\eeqas
%\end{widetext}
$(\b+1)_{\mu}$ is the Pochhammer symbol, $\ds\binom{m_1+a_1}{m_1}$ is the binomial term and $\a\in\mathbb{N}$.  $F_A^{(s)}\left(\ba{r} a:\overbrace{a_1,\dots,a_s}^{s}\\\underbrace{b_1,\dots,b_s}_{s}\ea;x_1,\dots,x_s\right)$ is the Lauricella hypergeometric function of $s$ variables $x_1,\dots,x_s$ and $2s+1$ parameters $a_1,\dots,a_s,b_1,\dots,b_s,a$ and it is defined by \cite{Srivastava}
%\begin{widetext}
\beq
F_A^{(s)}\left(\ba{r} a;\overbrace{a_1,\dots,a_s}^{s}\\\underbrace{b_1,\dots,b_s}_{s}\ea;x_1,\dots,x_s\right)=\ds\sum_{j_1,...,j_s=0}^{\infty}\f{a_{j_1+...+j_s}(a_1)_{j_1}...(a_s)_{j_s}}{(b_1)_{j_1}...(b_s)_{j_s}}\f{x_1^{j_1}...x_s^{j_s}}{j_1!...j_s!},
\eeq 
and $J_{2,(\ell,m)}^{(\a)}$ is the entropic moment of the rotational wave function $Y_{\ell,m}(\theta,\phi)$ and it is defined by \cite{jsd.ijqc2016} 
\beq\label{spherical.moment}
\ba{ll}
J_{2,(\ell,m)}^{(\a)}&=\ds\int_{\theta=0}^{\pi}\int_{\phi=0}^{2\pi} |Y_{\ell,m}(\theta,\phi)|^{2\a}\sin\theta\,d\theta\,d\phi\\\\
&=\ds\f{2^{2\a(2m-1)+2}(2\ell+1)^{\a}\G(m\a+1)^2}{\pi^{(2\a-1)}\G(2m\a+2)}\left[\f{\G(m+\f{1}{2})^2\G(m+1)^2\G(\ell-m+1)\G(\ell+m+1)}{\G(2m+1)^2\G(\ell+1)^2}\right]^{\a}B(\a,\ell,m),
\ea
\eeq 
where
\beq
B(\a,\ell,m)=\ds\binom{\ell}{\ell-m}^{2\a}\sum_{j_1,...,j_{2\a}=0}^{\ell-m}\f{(m\a+1)_{j_1+...+j_{2\a}}}{(2m\a+2)_{j_1+...+j_{2\a}}}\f{(m-\ell)_{j_1}(m+\ell+1)_{j_1}...(m-\ell)_{j_{2\a}}(m+\ell+1)_{j_{2\a}}}{(m+1)_{j_1}...(m+1)_{j_{2\a}}\,j_1!...j_{2\a}!},
\eeq
$\bar{\mu}_1=\a L+\f{1}{2}$, and $C=\f{a^{\f{3\a-1}{2}}2^{\a-1}(n!)^{\a}}{a^{\bar{\mu}_1+1}\left[\G(L+\f{3}{2})\right]^{\a}}$. R\'enyi entropy and information theoretic measure of Laguerre polynomial is addressed in refs. \cite{pshop9,jsd.Laguerre,jsd.ijqc2011,jsd.amc}. 
\begin{figure}[h] % Fig 1 Renyi entropy 0,0,0 \lam 
	\centering
	\includegraphics[width=12cm,height=8cm]{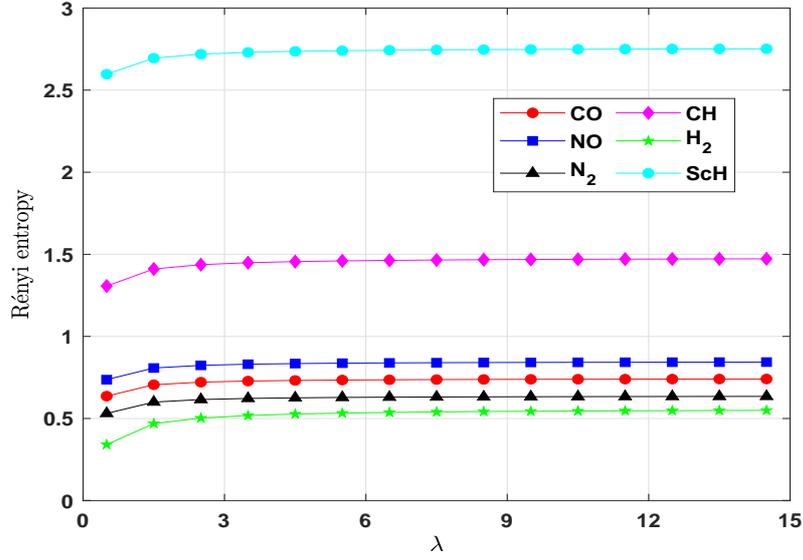}	
	\caption{\label{Fig1} Plot of R\'enyi entropies of some diatomic molecules of state $\ds\widehat{\psi}_{0,0,0}({\bf r},\lam)$ with respect to $\lam$ for $\a=2.5$.}
\end{figure}
Similarly, R\'enyi entropy of (\ref{rho.iso.n}) is defined by \cite{pshop9,jsd.Laguerre,jsd.ijqc2011,jsd.amc}
\beq
\ba{l}%gin{eqnarray}
\widehat{\mathcal{R}}_{n,\ell,m}^{(\a)}(\lam)=\f{1}{1-\a}\ln\left[\ds\f{(2a\sqrt{a})^{\a-1}(\lam+1)^{\a}}{\left[\lam\G(L+\f{3}{2})\right]^{\a}}
\sum\limits_{\substack{j,p=0 \\ k_1+...+k_j=p}}^{\infty}\f{(-1)^{j+p}\binom{2\a+j-1}{j}(\G(\bar{\mu}_2+1)\,J_{2,(\ell,m)}^{(\a)})}{\left[\lam\,\G(L+\f{3}{2})\right]^j\a^{\bar{\mu}_2+1}k_1!...k_j!(L+\f{3}{2}+k_1)...(L+\f{3}{2}+k_j)}
\right],n=0,
\ea
\eeq
\beq%gin{widetext}
\ba{l}%egin{eqnarray}
\widehat{\mathcal{R}}_{n,\ell,m}^{(\a)}(\lam)=\f{1}{1-\a}\ln\left[\f{(2a\sqrt{a})^{\a-1}(n!)^{\a}}{\left[\G(L+\f{3}{2})\right]^{\a}}\ds\sum_{i=0}^{2\a}\sum\limits_{\substack{j,p=0 \\ k_1+...+k_j=p}}^{\infty}\f{(-1)^{i+j+p}\binom{2\a}{i}\binom{i+j-1}{j}}{n^i\left[\lam\,\G(L+\f{3}{2})\right]^{i+j}}\right.\\
\left.~~~\times\ds\f{\ds A_0\left(\bar{\mu}_3,0,2\a,\left\{\overbrace{n-1}^i,\overbrace{n}^{2\a-i}\right\},\left\{\overbrace{L+\f{3}{2}}^i,\overbrace{L+\f{1}{2}}^{2\a-i}\right\},\left\{\overbrace{\f{1}{\a+i}}^{2\a}\right\}\right)}{(\a+i)^{\bar{\mu}_3+1}k_1!...k_j!(L+\f{3}{2}+k_1)...(L+\f{3}{2}+k_j)}J_{2,(\ell,m)}^{(\a)}\right],~n=1,2,\dots,
\ea
\eeq
where 
\beq
\ba{r}\label{thetabar.G}
A_p\left(\mu_3,\b,2\a,\left\{\overbrace{n-1}^i,\overbrace{n}^{2\a-i}\right\},\left\{\overbrace{L+\f{3}{2}}^i,\overbrace{L+\f{1}{2}}^{2\a-i}\right\},\left\{\overbrace{\f{1}{\a+i}}^{2\a}\right\}\right)=\ds(\b+1)_{\mu_3}\binom{n+L+\f{1}{2}}{n-1}^i\binom{n+L+\f{1}{2}}{n}^{2\a-i}\\\\
\ds\times F_A^{(2\a+1)}\left(\ba{r}\mu_3+\b+1;\overbrace{-n+1,...,-n+1}^{i},\overbrace{-n,...,-n}^{2\a-i},-p\\\underbrace{L+\f{5}{2},...,L+\f{5}{2}}_{i},\underbrace{L+\f{3}{2},...,L+\f{3}{2}}_{2\a-i},\beta+1\ea;\overbrace{\f{1}{\a+i},...,\f{1}{\a+i}}^{2\a},1\right),
\ea 
\eeq 
%\end{widetext}
and
\beq
\ba{l}
\bar{\mu}_2=p+(L+\f{3}{2})j+\a L+\f{1}{2},\\
\bar{\mu}_3=p+(L+\f{3}{2})(i+j)+\a L+\f{1}{2},
\ea
\eeq
$\a\in\mathbb{N}$ and $\lam\in(-\infty,-2)\cup(1,\infty)$. 
\subsection{R\'enyi complexity ratio}
\begin{figure}[h] % Fig 2 RCR wrt n
	\centering
	\includegraphics[width=12cm,height=8cm]{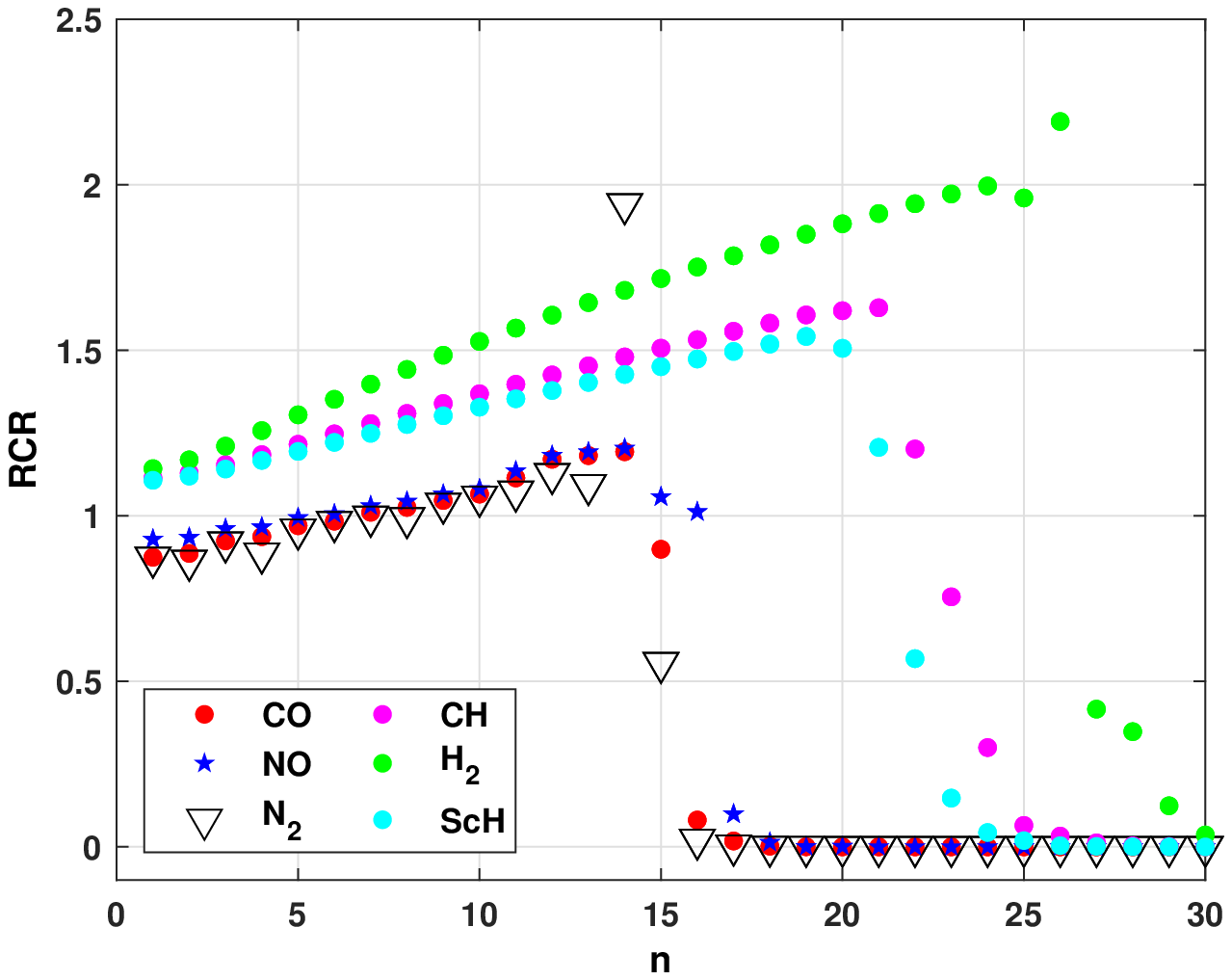}%~	
	\caption{\label{Fig2} Compare R\'enyi complexity ratios of some diatomic molecules between $\ds\left(\widehat{\psi}_{n,0,0}({\bf r},\lam),\psi_{n,0,0}({\bf r})\right)$ with respect to $n$ of order $(\a,\b)=(2.25,3.5)$ and $\lam=2.5$.}
\end{figure}

\begin{figure}[h] % Fig 3 RCR D_e, r_e, m_{\mu} 
	\centering
	\includegraphics[width=18cm,height=12cm]{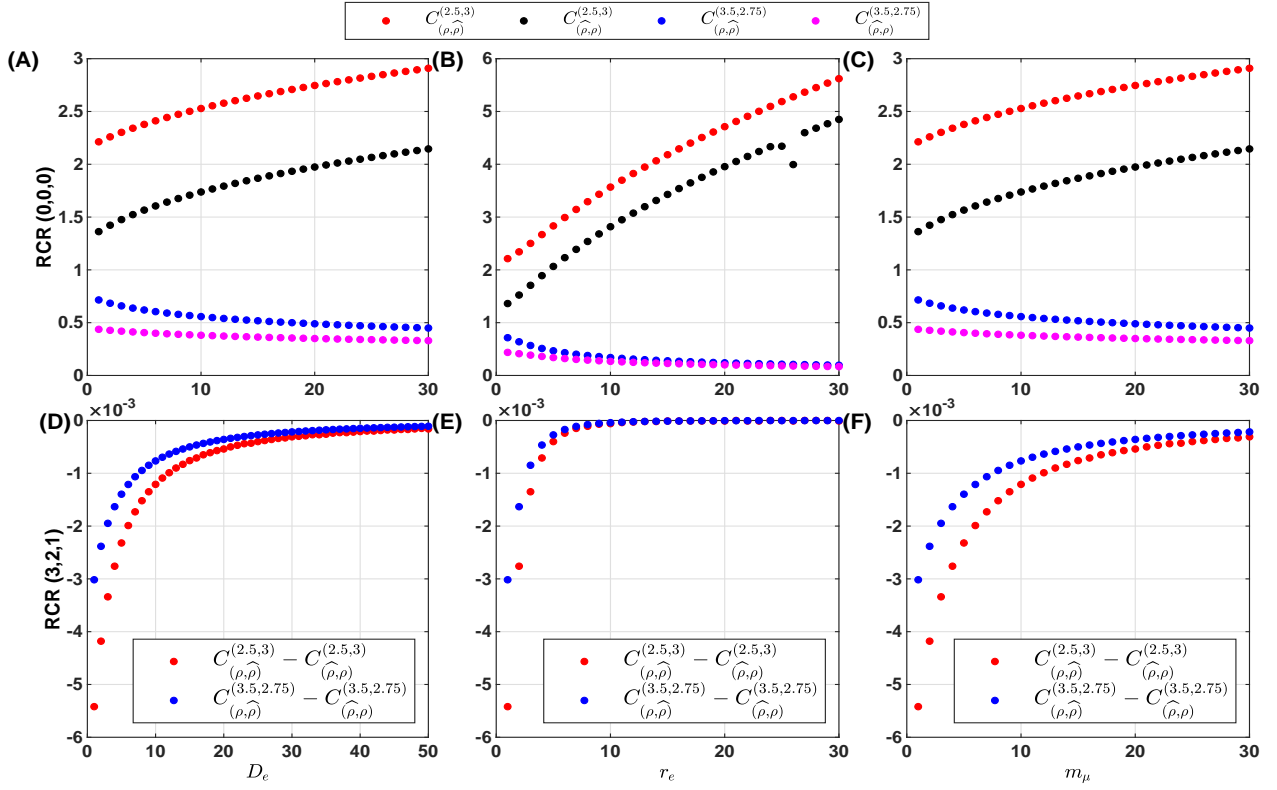}	
	\caption{\label{Fig3} Compare RCR with respect to (A),(D) $D_e$ for $r_e=m_{\mu}=1$ (B),(E) $r_e$ for $D_e=m_{\mu}=1$, (C),(F) $m_{\mu}$ for $D_e=r_e=1$. The others parameters are $\lam=1.5$, $\hbar=1$, for first row $(n,\ell,m)=(0,0,0)$ and for second row $(n,\ell,m)=(3,2,1)$. The red (dashed), black (dotted), blue (dashed) and magenta (dotted) curves represnt $C_{(\rho,\widehat{\rho})}^{(2.5,3)}$, $C_{(\widehat{\rho},\rho)}^{(2.5,3)}$, $C_{(\rho,\widehat{\rho})}^{(3.5,2.75)}$ and $C_{(\widehat{\rho},\rho)}^{(3.5,2.75)}$ respectively.}
\end{figure}
The explicit form of exact value of the R\'enyi complexity ratio of $\widehat{\rho}_{n,\ell,m}$ and $\rho_{n,\ell,m}$ for positive integral order is defined by
%\begin{widetext}
\beq\label{RCR.0}
\ba{l}
C^{(\a,\b)}_{(\widehat{\rho},\rho),n,\ell,m}=\left\{\ba{l}\left(\f{\lam+1}{\lam}\right)^{\f{\a}{1-\a}}\left(\f{2\sqrt{a}(2\ell+1)\G(m+\f{1}{2})^2\G(m+1)^2\G(\ell-m+1)\G(\ell+m+1)}{\G(L+\f{3}{2})\G(2m+1)^2\G(\ell+1)^2}\right)^{\f{\a}{1-\a}-\f{2\b}{1-\b}}\\\\
\times \f{2^{\f{\a(4m-2)+1}{1-\a}-\f{\b(4m-2)+1}{1-\b}}\b^{\f{\bar{\mu}'_1+1}{1-\b}}}{\left[A_0\left(\bar{\mu}'_1,0,2\b,\{n\},\left\{L+\f{1}{2}\right\},\left\{\f{1}{\b}\right\}\right)\right]^{\f{1}{1-\b}}}\f{a^{\f{2\a-3}{2-2\a}-\f{2\b-1}{2-2\b}}}{\pi^{\f{2\a-1}{1-\a}-\f{2\b-1}{1-\b}}}\f{\left[\G(m\a+1)\right]^{\f{2\a}{1-\a}}}{\left[\G(m\b+1)\right]^{\f{2\b}{1-\b}}} \ds\f{\left[\G(2m\b+2)\right]^{\f{1}{1-\b}}}{\left[\G(2m\a+2)\right]^{\f{1}{1-\a}}}\f{\left[\ds B(\a,\ell,m)\right]^{\f{1}{1-\a}}}{\left[\ds B(\b,\ell,m)\right]^{\f{1}{1-\b}}}\\\\
\times\left(\ds\sum\limits_{\substack{j,p=0 \\ k_1+...+k_j=p}}^{\infty}\f{(-1)^{j+p}\binom{2\a+j-1}{j}}{\left[\lam\,\G(L+\f{3}{2})\right]^j}\ds\f{\ds\G(\bar{\mu}_2+1)}{\a^{\bar{\mu}_2+1}k_1!...k_j!(L+\f{3}{2}+k_1)...(L+\f{3}{2}+k_j)}\right)^{\f{1}{1-\a}}\ea\right\},n=0,
\ea
\eeq

\beq\label{RCR.n}
\ba{l}
\small C^{(\a,\b)}_{(\widehat{\rho},\rho),n,\ell,m}=
\left\{\ba{l}\ds\f{a^{\f{2\a-3}{2-2\a}-\f{2\b-1}{2-2\b}}}{\pi^{\f{2\a-1}{1-\a}-\f{2\b-1}{1-\b}}}\left(\f{2\sqrt{a}(n!)(2\ell+1)\G(m+\f{1}{2})^2\G(m+1)^2\G(\ell-m+1)\G(\ell+m+1)}{\G(n+L+\f{3}{2})\G(2m+1)^2\G(\ell+1)^2}\right)^{\f{\a}{1-\a}-\f{\b}{1-\b}}\\\\
\ds\f{2^{\f{2\a(2m-1)+1}{1-\a}-\f{2\b(2m-1)+1}{1-\b}}}{\left[\f{1}{\b^{\bar{\mu}'_1+1}}A_0\left(\bar{\mu}'_1,0,2\b,\{n\},\left\{L+\f{1}{2}\right\},\left\{\f{1}{\b}\right\}\right)\right]^{\f{1}{1-\b}}}\f{\left[\G(m\a+1)\right]^{\f{2\a}{1-\a}}}{\left[\G(m\b+1)\right]^{\f{2\b}{1-\b}}}\ds\f{\left[\G(2m\b+2)\right]^{\f{1}{1-\b}}}{\left[\G(2m\a+2)\right]^{\f{1}{1-\a}}}\f{\left[\ds B(\a,\ell,m)\right]^{\f{1}{1-\a}}}{\left[\ds B(\b,\ell,m)\right]^{\f{1}{1-\b}}}\\\\
\tiny{\left(\sum_{i=0}^{2\a}\sum\limits_{\substack{j,p=0 \\ k_1+...+k_j=p}}^{\infty} \f{(-1)^{i+j+p}\binom{2\a}{i}\binom{i+j-1}{j}}{n^i\left[\lam\,\G(L+\f{3}{2})\right]^{i+j}}\f{A_0\left(\bar{\mu}_3,0,2\a,\left\{\overbrace{n-1}^i,\overbrace{n}^{2\a-i}\right\},\left\{\overbrace{L+\f{3}{2}}^i,\overbrace{L+\f{1}{2}}^{2\a-i}\right\},\left\{\overbrace{\f{1}{\a+i}}^{2\a}\right\}\right)}{(\a+i)^{\bar{\mu}_3+1}k_1!...k_j!(L+\f{3}{2}+k_1)...(L+\f{3}{2}+k_j)}\right)^{\f{1}{1-\a}}}\ea\right\},\\
~~~~n=1,2,3,\dots 
\ea
\eeq
%\end{widetext}
where $\bar{\mu}'_1=\b L+\f{1}{2}$, $\a,\b\in\mathbb{N}$ and $\lam\in (-\infty,-2)\cup(1,\infty)$. Similarly, one can find the exact value of $C_{(\rho,\widehat{\rho}),n,\ell,m}^{(\a,\b)}$. 
\begin{figure}[h] % Fig 4 RCR 0,0,0 
	\centering
	\includegraphics[width=18cm,height=12cm]{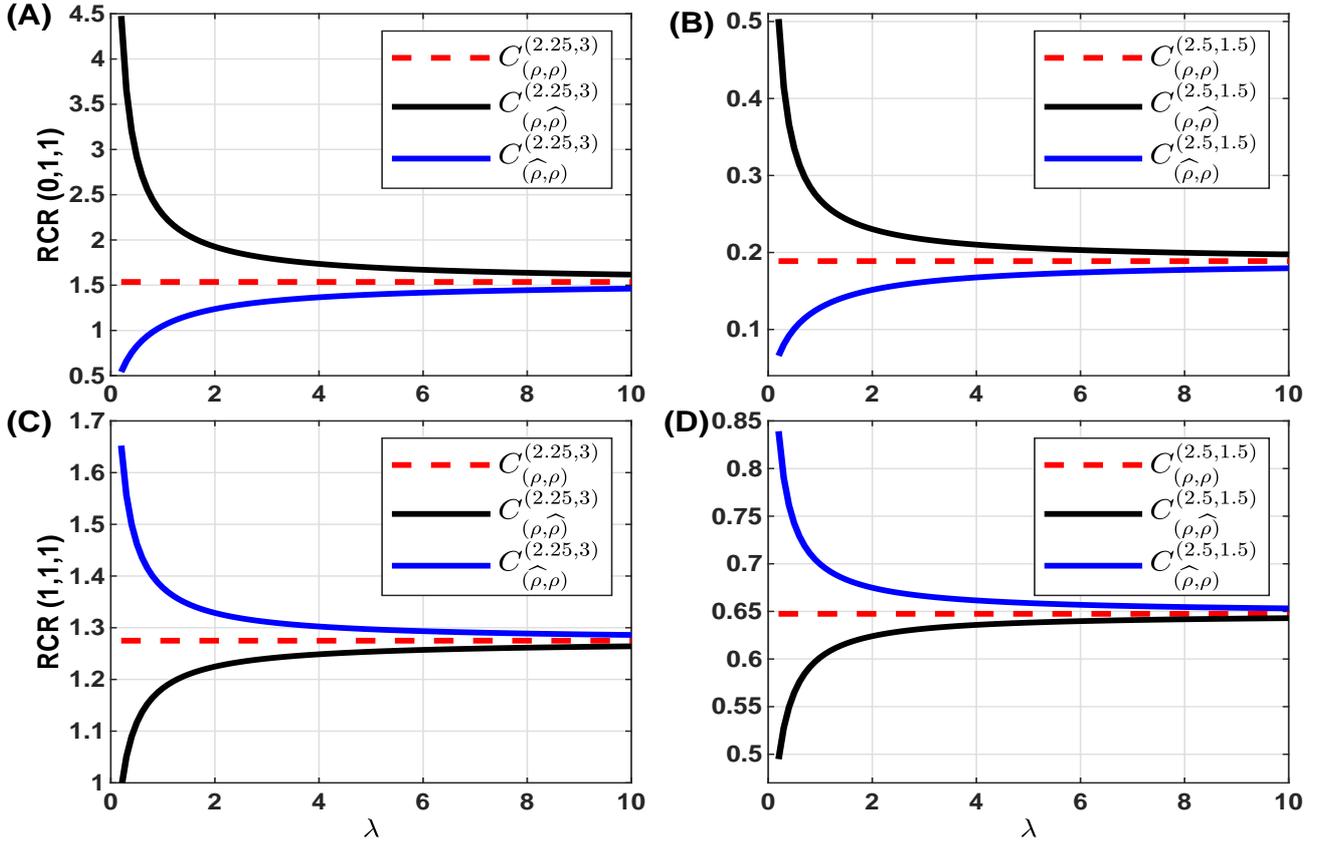}	
	\caption{\label{Fig4} Plot of RCR with respect to $\lam$ for $D_e=\f{7}{2},r_e=\f{1}{2},m_{\mu}=1,\hbar=1$, (A)-(B) $(n,\ell,m)=(0,1,1)$ and (C)-(D) $(n,\ell,m)=(1,1,1)$.}
\end{figure}

\subsection{Generalized R\'enyi complexity and shape R\'enyi complexity }
\begin{figure}[h] % Fig 5 GRC Iso with GRC rho 0,0,0 
	\centering
	\includegraphics[width=18cm,height=12cm]{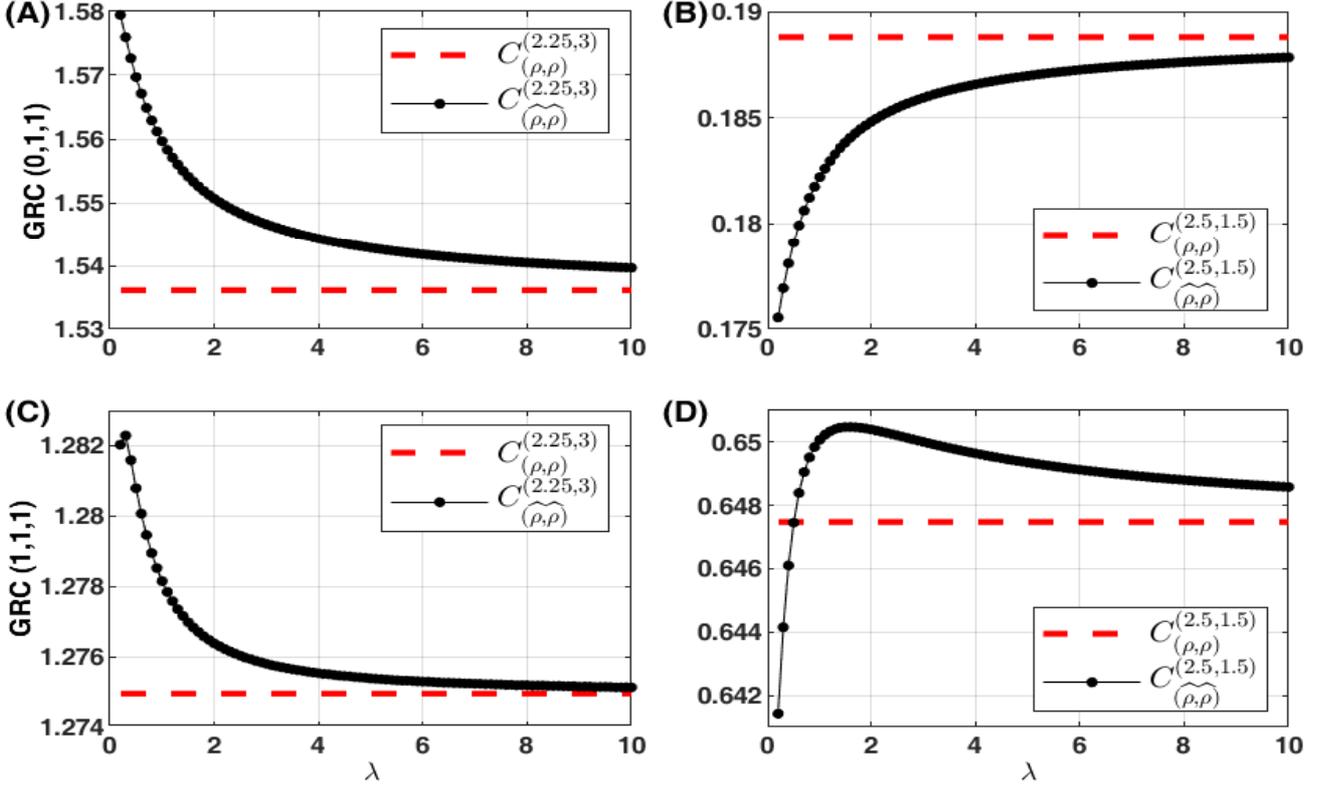}	
	\caption{\label{Fig5} Plot of GRC with respect to $\lam$ for $D_e=\f{7}{2},r_e=\f{1}{2},m_{\mu}=1,\hbar=1$, (A)-(B) $(n,\ell,m)=(0,1,1)$ and (C)-(D) $(n,\ell,m)=(1,1,1)$.}
\end{figure}

\begin{figure}[h] % Fig 6 GRC 0,0,0 1
	\centering
	\includegraphics[width=12cm,height=8cm]{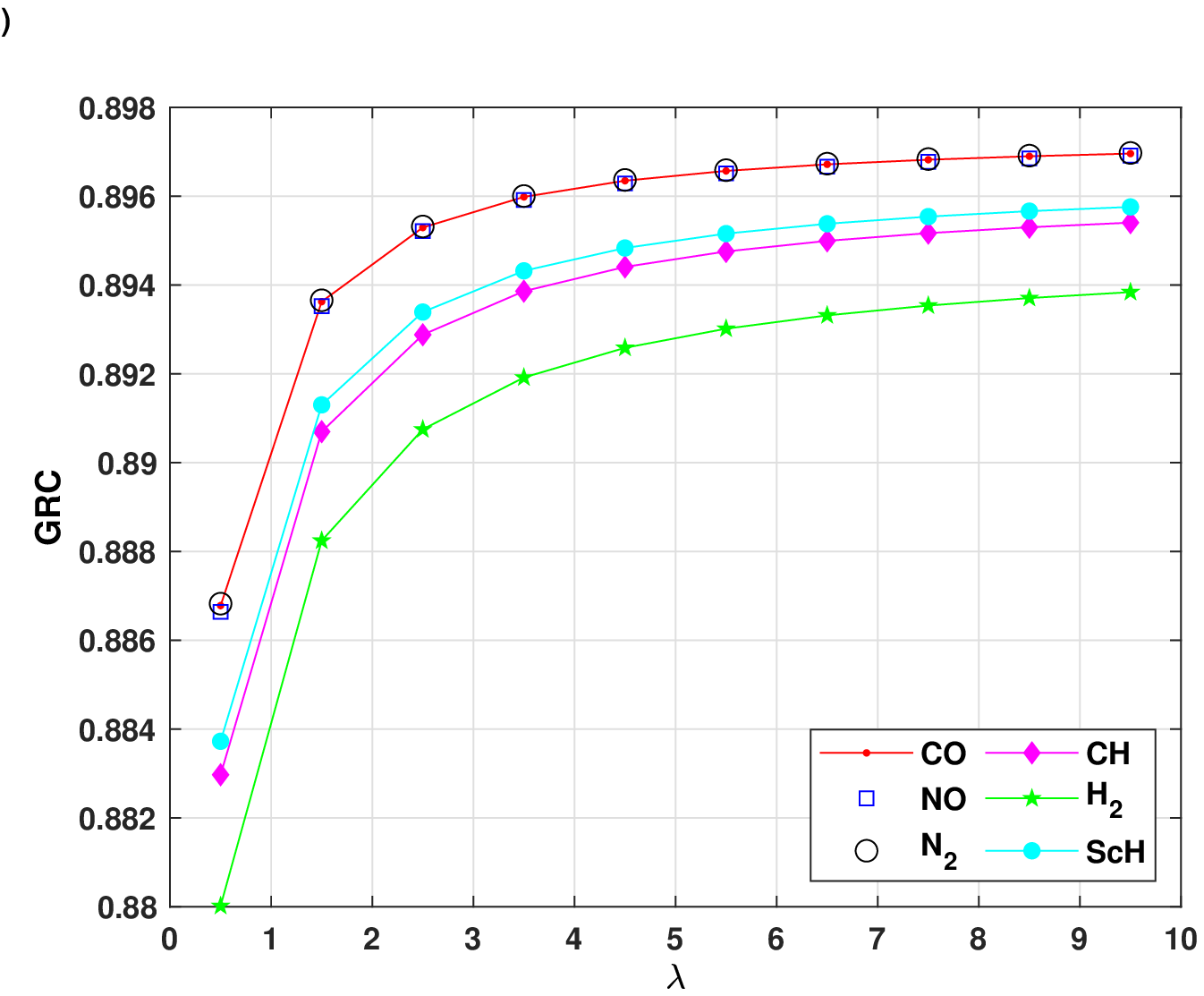}
	\caption{\label{Fig6} Plot of generalized R\'enyi complexities of some diatomic molecules of the state $\ds\widehat{\psi}_{0,0,0}({\bf r},\lam)$ with respect to $\lam$ for $(\a,\b)=(8.5,3.5)$.}
\end{figure}
The explicit form of generalized R\'enyi complexity of $\psi_{n,\ell,m}({\bf r})$ and $\widehat{\psi}_{n,\ell,m}({\bf r},\lam)$ are respectively defined by
%\begin{widetext}
\beq
\ba{l}\label{GRC.rho}
C_{n,\ell,m}^{(\a,\b)}=\ds\left(\f{2\sqrt{a}(n!)}{\G(L+\f{3}{2})}\right)^{\f{\a}{1-\a}-\f{\b}{1-\b}}\f{\b^{\f{\bar{\mu}'_1+1}{1-\b}}}{\a^{\f{\bar{\mu}_1+1}{1-\a}}}\f{\left[\f{\sqrt{a}}{2}\,A_0\left(\bar{\mu}_1,0,2\a,\{n\},\left\{L+\f{1}{2}\right\},\left\{\f{1}{\a}\right\}\right)J^{(a)}_{2,(\ell,m)}\right]^{\f{1}{1-\a}}}{\left[\f{\sqrt{a}}{2}\,A_0\left(\bar{\mu}'_1,0,2\b,\{n\},\left\{L+\f{1}{2}\right\},\left\{\f{1}{\b}\right\}\right)J_{2,(\ell,m)}^{(\b)}\right]^{\f{1}{1-\b}}},~n=0,1,2,\dots,
\ea
\eeq 
and  
\beq\label{GRC.rho.iso.0}
\ba{l}
\widehat{C}_{n,\ell,m}^{(\a,\b)}(\lam)=\ds\f{\left(\f{2\sqrt{a}(\lam+1)}{\lam\G(L+\f{3}{2})}\right)^{\f{\a}{1-\a}-\f{\b}{1-\b}}\left(\ds\sum\limits_{\substack{j,p=0 \\ k_1+...+k_j=p}}^{\infty}\f{(-1)^{j+p}\binom{2\a+j-1}{j}\G(\bar{\mu}_2+1)\,J_{2,(\ell,m)}^{(\a)}}{\left[\lam\,\G(L+\f{3}{2})\right]^j\a^{\bar{\mu}_2+1}k_1!...k_j!(L+\f{3}{2}+k_1)...(L+\f{3}{2}+k_j)}\right)^{\f{1}{1-\a}}}{\left(\f{1}{2\sqrt{a}}\right)^{\f{1}{1-\b}-\f{1}{1-\a}}\left(\ds\sum\limits_{\substack{j,p=0 \\ k_1+...+k_j=p}}^{\infty}\f{(-1)^{j+p}\binom{2\b+j-1}{j}\G(\bar{\mu}'_2+1)\,J_{2,(\ell,m)}^{(\b)}}{\left[\lam\,\G(L+\f{3}{2})\right]^j\b^{\bar{\mu}'_2+1}k_1!...k_j!(L+\f{3}{2}+k_1)...(L+\f{3}{2}+k_j)}\right)^{\f{1}{1-\b}}},n=0,
\ea
\eeq
\beq\label{GRC.rho.iso.n}
\ba{l}
\widehat{C}_{n,\ell,m}^{(\a,\b)}(\lam)=\ds\left(\f{2\sqrt{a}(n!)}{\G(L+\f{3}{2})}\right)^{\f{\a}{1-\a}-\f{\b}{1-\b}}\,\f{J_{2,(\ell,m)}^{(\a)}}{J_{2,(\ell,m)}^{(\a)}}\\
\times \tiny{\ds\f{\left(\ds\sum_{i=0}^{2\a}\sum\limits_{\substack{j,p=0 \\ k_1+...+k_j=p}}^{\infty}\f{(-1)^{i+j+p}\binom{2\a}{i}\binom{i+j-1}{j}}{2\sqrt{a}\,n^i\left[\lam\G(L+\f{3}{2})\right]^{i+j}}\f{A_0\left(\bar{\mu}_3,0,2\a,\left\{\overbrace{n-1}^i,\overbrace{n}^{2\a-i}\right\},\left\{\overbrace{L+\f{3}{2}}^i,\overbrace{L+\f{1}{2}}^{2\a-i}\right\},\left\{\overbrace{\f{1}{\a+i}}^{2\a}\right\}\right)}{(\a+i)^{\bar{\mu}_3+1}k_1!...k_j!(L+\f{3}{2}+k_1)...(L+\f{3}{2}+k_j)}\right)^{\f{1}{1-\a}}}{\left(\ds\sum_{i=0}^{2\b}\sum\limits_{\substack{j,p=0 \\ k_1+...+k_j=p}}^{\infty}\f{(-1)^{i+j+p}\binom{2\b}{i}\binom{i+j-1}{j}}{2\sqrt{a}\,n^i\left[\lam\G(L+\f{3}{2})\right]^{i+j}}\f{A_0\left(\bar{\mu}'_3,0,2\b,\left\{\overbrace{n-1}^i,\overbrace{n}^{2\a-i}\right\},\left\{\overbrace{L+\f{3}{2}}^i,\overbrace{L+\f{1}{2}}^{2\a-i}\right\},\left\{\overbrace{\f{1}{\a+i}}^{2\a}\right\}\right)}{(\b+i)^{\bar{\mu}'_3+1}k_1!...k_j!(L+\f{3}{2}+k_1)...(L+\f{3}{2}+k_j)}\right)^{\f{1}{1-\b}}}},
 ~~~n=1,2,3,\dots
\ea
\eeq
%\end{widetext}
where
\beq
\ba{l}
\bar{\mu}'_2=p+(L+\f{3}{2})j+\b L+\f{1}{2}\\
\bar{\mu}'_3=p+(L+\f{3}{2})(i+j)+\b L+\f{1}{2}
\ea
\eeq
Similarly, one can find the exact values of SRC of $\psi_{n,\ell,m}({\bf r})$ and $\widehat{\psi}_{n,\ell,m}({\bf r},\lam)$ from Eqs. (\ref{GRC.rho}), (\ref{GRC.rho.iso.0}) and (\ref{GRC.rho.iso.n}), replacing $\b$ by two. 

\begin{figure}[h] % Fig 7 GRC 0,0,0 2
	\centering
	\includegraphics[width=12cm,height=8cm]{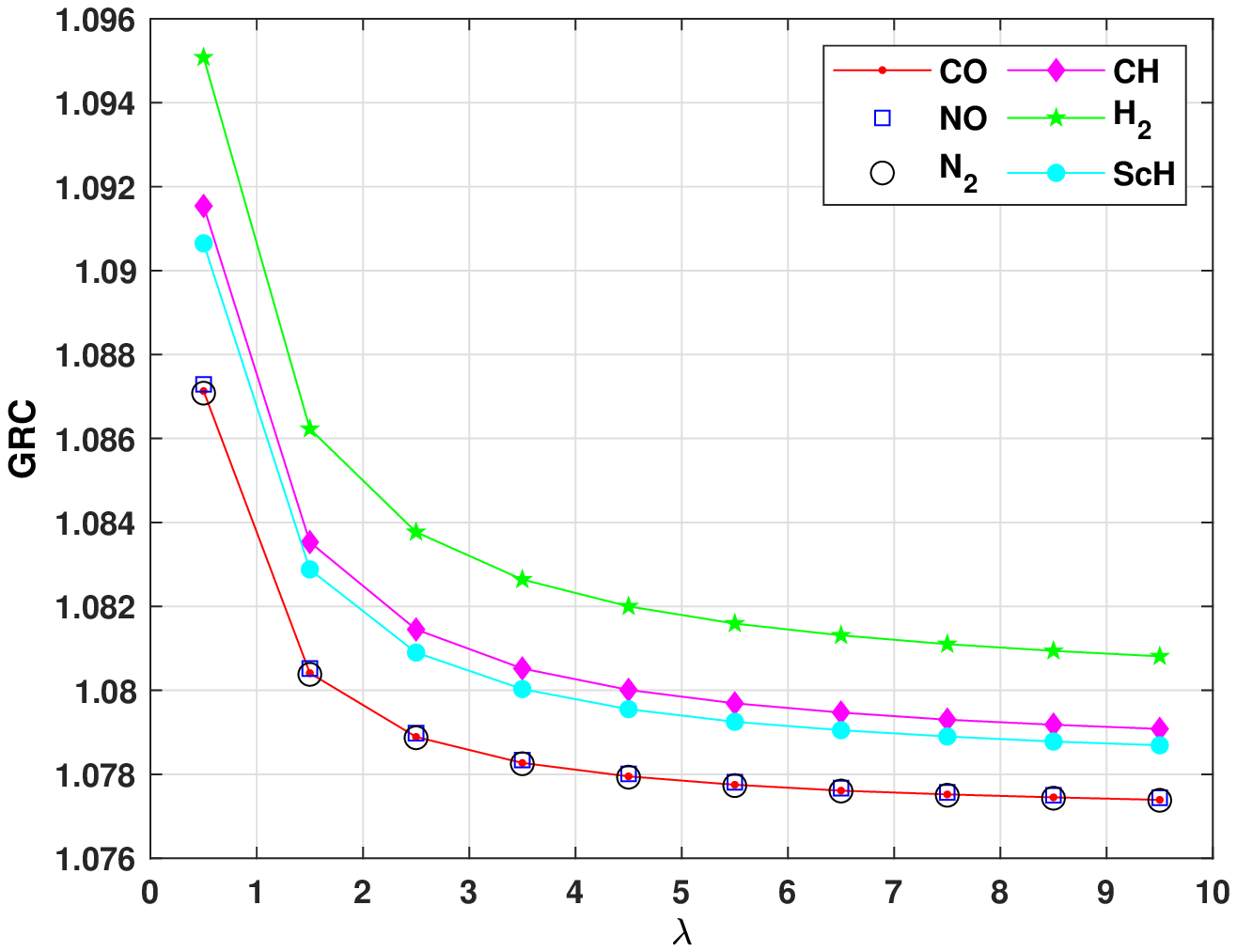}%~	
	\caption{\label{Fig7} Plot of generalized R\'enyi complexities of some diatomic molecules of the state $\ds\widehat{\psi}_{0,0,0}({\bf r},\lam)$ with respect to $\lam$ for $(\a,\b)=(2.25,3.5)$.}
\end{figure}

\begin{figure}[h] % Fig 8 SRC 0,0,0 1
	\centering
	\includegraphics[width=12cm,height=8cm]{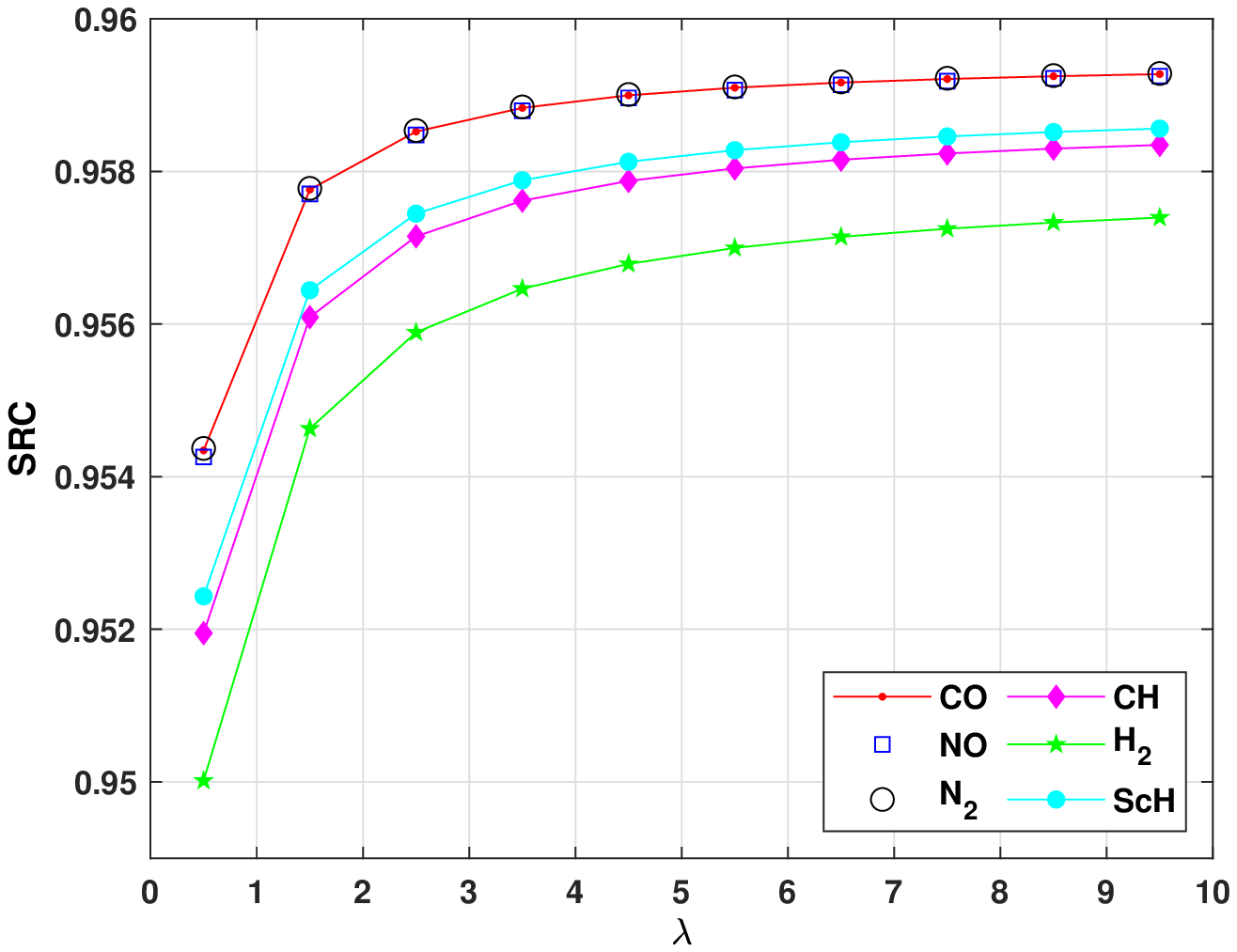}%~	
	\caption{\label{Fig8} Plot of shape R\'enyi complexities of some diatomic molecules of the state $\ds\widehat{\psi}_{0,0,0}({\bf r},\lam)$ with respect to $\lam$ for $\a=2.5$.}
\end{figure}

\begin{figure}[h] % Fig 9 SRC 0,0,0 2
	\centering
	\includegraphics[width=12cm,height=8cm]{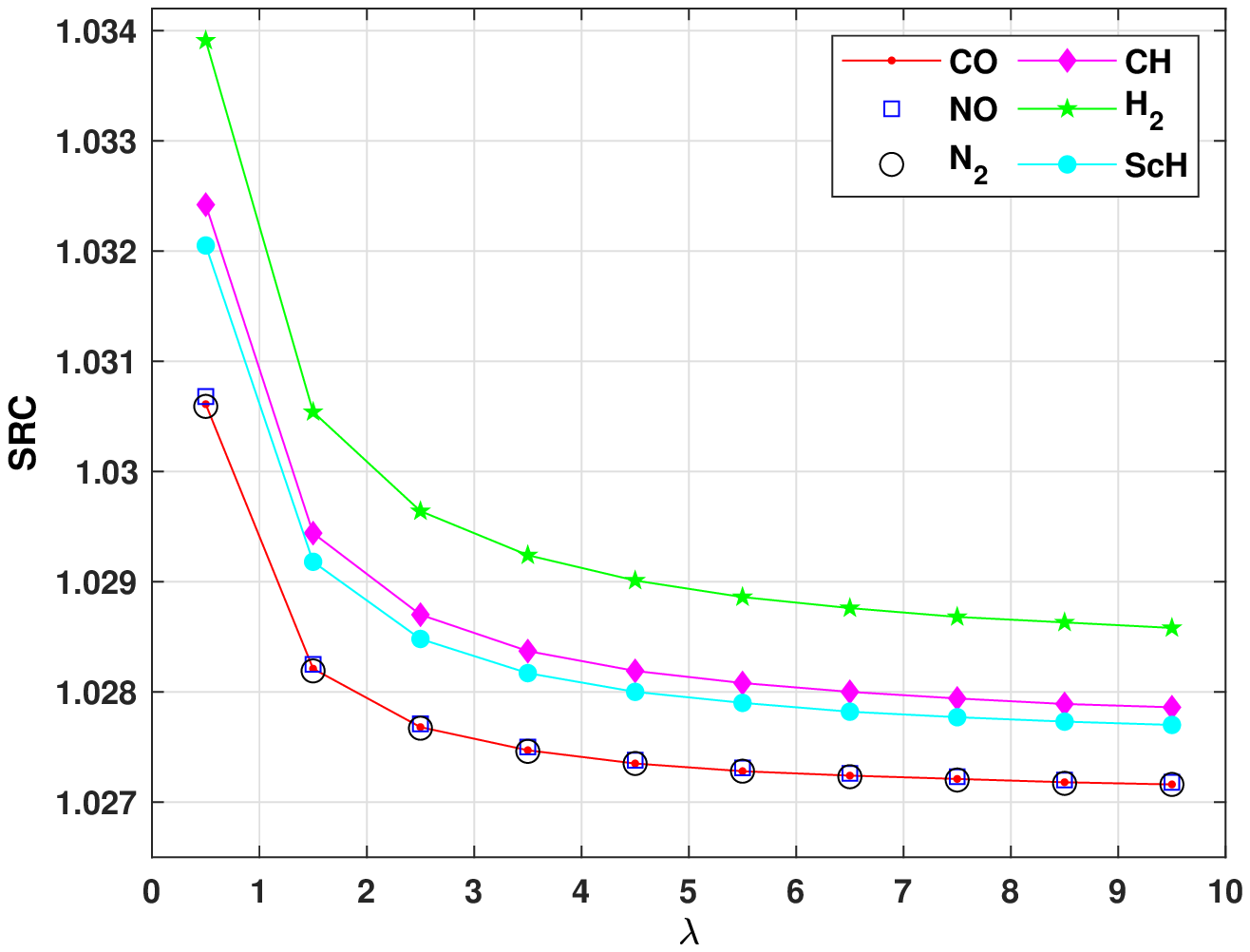}%~	
	\caption{\label{Fig9} Plot of shape R\'enyi complexities of some diatomic molecules of the state $\ds\widehat{\psi}_{0,0,0}({\bf r},\lam)$ with respect to $\lam$ for $\a=1.75$.}
\end{figure}
\begin{figure}[h] % Fig 9 SRC 0,0,0 2
	\centering
	\includegraphics[width=18cm,height=12cm]{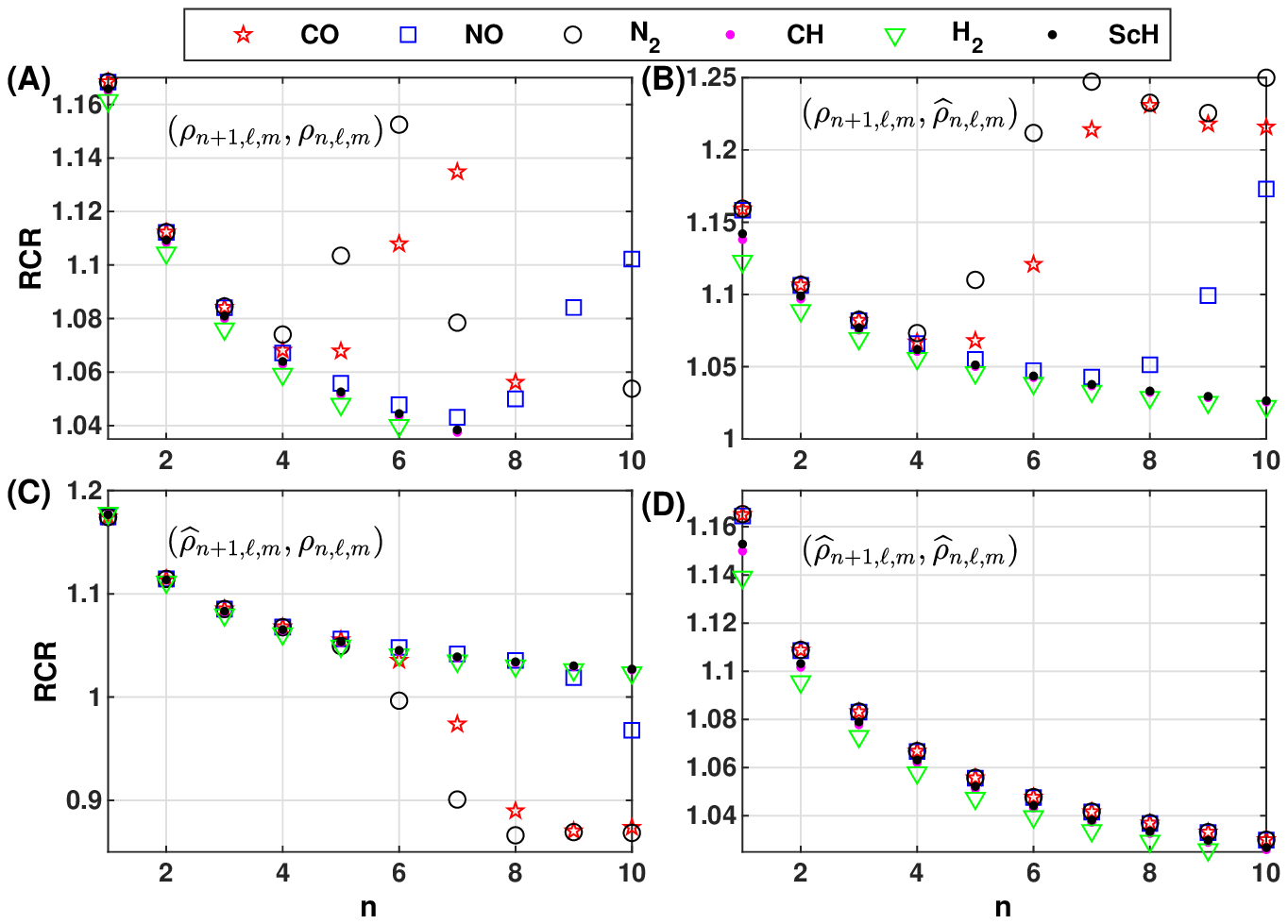}%~	
	\caption{\label{Fig10} Plot of R\'enyi complexity ratios of some diatomic molecules between two succesive energy levels $(n+1,0,m)$ and $(n,0,m)$ for $\lam=2.5$, $(\a,\b)=(2.5,2.5)$.}
\end{figure}
\begin{figure}[h] % Fig 9 SRC 0,0,0 2
	\centering
	\includegraphics[width=18cm,height=12cm]{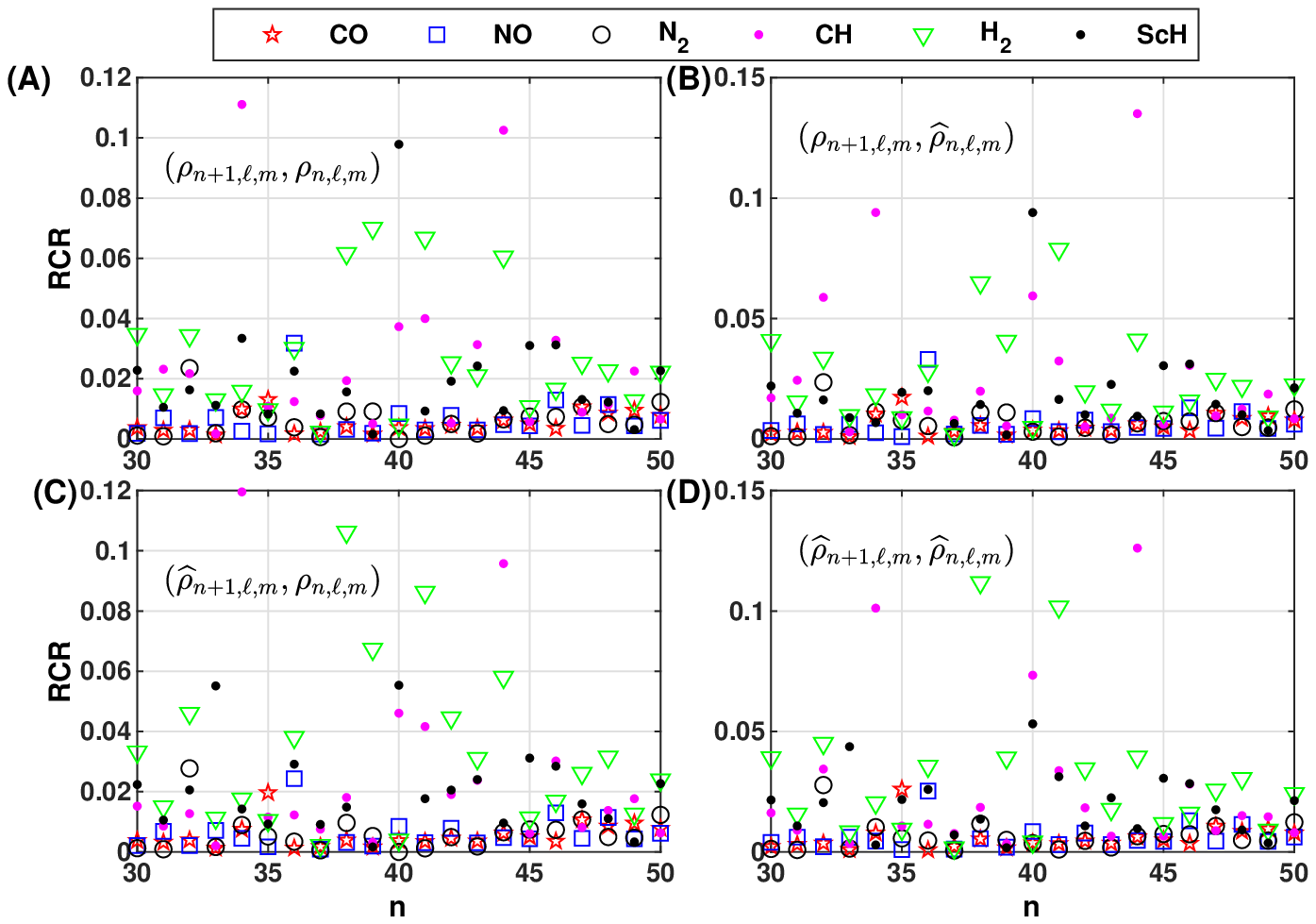}%~	
	\caption{\label{Fig11} Plot of R\'enyi complexity ratios of some diatomic molecules between two succesive energy levels $(n+1,0,m)$ and $(n,0,m)$ for $\lam=2.5$, $(\a,\b)=(2.5,2.5)$.}
\end{figure}
\section{Results and discussions}\label{results}
The total density function of a state with quantum numbers $(n,\ell,m)$ is separable for the pseudoharmonic oscillator potential for $n=0,1,2,\cdots$; $\ell=0,1,2,\cdots$; and $m=0,\pm1,\pm2,\dots,\pm\ell$. Under this circumstances the pseudoharmonic potential and a family of its isospectral potentials describe exact forms of R\'enyi entropy, RCR, GRC and SRC. The harmonic spherical function $Y_{\ell,m}$ is defined in a closed and bounded domain $[0,\pi]\times[0,2\pi]$ and the radial wave functions $\sqrt{a}N_{n}e^{-\f{1}{2}ar^2}\,(\sqrt{a}r)^{L}\ds L_{n}^{L+\f{1}{2}}\left(ar^2\right)$  and $\widehat{C}_n\sqrt{\Om(r,\lam)}\Phi_{n}(r)$ are bounded in $[0,\infty)$. Then the corresponding total density functions have effective domains in $[0,\infty)\times[0,\pi]\times[0,2\pi]$. The energy level spacing of pseudoharmonic and its isospectral potentials are same and it describes internuclear potential-energy function of diatomic molecules \cite{pho.import} but wave functions for isospectral potentials do not match with diatomic molecules. In this section, we will find the numerical values of R\'enyi entropy, RCR, GRC and SRC of solutions (\ref{sol.pseudo}) and (\ref{sol.pseudo.iso}). 

%We observe that $\widehat{\psi}_{0,\ell,m}$ is more affected by $\lam$, and $|\psi_{0,\ell,m}({\bf r})-\widehat{\psi}_{0,\ell,m}({\bf r},\lam)|\ge |\psi_{n,\ell,m}({\bf r})-\widehat{\psi}_{n,\ell,m}({\bf r},\lam)|$, for $n=1,2,\cdots$, ${\bf r}\in [0,\infty)\times[0,\pi]\times[0,2\pi]$ and admisible value of $\lam$ .

The numerical values of R\'enyi entropies of $\widehat{\psi}_{0,0,0}({\bf r},\lam)$ of order $\a=2.5$ for CO, NO, N$_2$, CH, H$_2$, and ScH diatomic molecules are plotted in Fig. \ref{Fig1} with respect to $\lam$. The molecular parameter values of $D_e,r_e,\mu$ are taken from refs. \cite{data,pshop6.kds.oyewumi.jmc2012,DN.IJQC.Renyi} and we have considered $1~ amu=931.494028~ MeV/c^2$, $1~ cm^{-1} = 1.239841875\times 10^{-4}~ eV$, \&~ $c\hbar=1973.29~ eV\AA$, where $c$ is the speed of light. Note that R\'enyi entropies increase and go to some fixed numbers as $\lam$ increases.

Figs. \ref{Fig2} shows RCR of $\ds\left(\widehat{\psi}_{n,0,0}({\bf r},\lam),\psi_{n,0,0}({\bf r})\right)$ of rational order $(\a,\b)=(2.25,3.5)$ for same molecules with respect to $n$ for $\lam=2.5$. From this figure, one can see that, RCR goes to zero as $n$ increases for these selected molecules. Similarly, numerical values of RCR's of rational pair orders such as $C_{(\rho,\widehat{\rho})}^{(2.5,3)}$, $C_{(\widehat{\rho},\rho)}^{(2.5,3)}$, $C_{(\rho,\widehat{\rho})}^{(3.5,2.75)}$ and $C_{(\widehat{\rho},\rho)}^{(3.5,2.75)}$ are plotted with respect to $D_e,r_e,m_{\mu}$ in Fig. \ref{Fig3}, for first row $(n,\ell,m)=(0,0,0)$ and for second row $(n,\ell,m)=(3,2,1)$. Similarly, RCR's $C_{(\rho,\widehat{\rho})}^{(2.25,3)}$, $C_{(\widehat{\rho},\rho)}^{(2.25,3)}$, $C_{(\rho,\widehat{\rho})}^{(2.5,1.5)}$ and $C_{(\widehat{\rho},\rho)}^{(2.5,1.5)}$ are plotted with respect to $\lam$  in Fig. \ref{Fig4} for first row $(n,\ell,m)=(0,1,1)$ and for the second row $(n,\ell,m)=(1,1,1)$. All curves of Figs. \ref{Fig3} and \ref{Fig4} are plotted for some atomic values of parameters. From Figs. \ref{Fig3} and \ref{Fig4} it is clear that RCR's are monotone functions of $D_e,r_e,m_{\mu},|\lam|$. Moreover,  from Fig. \ref{Fig4} we observe that RCR's $C_{(\rho,\widehat{\rho})}^{(2.25,3)},~C_{(\widehat{\rho},\rho)}^{(2.25,3)}\rightarrow C_{(\rho,\rho)}^{(2.25,3)}$ and $C_{(\rho,\widehat{\rho})}^{(2.5,1.5)},~C_{(\widehat{\rho},\rho)}^{(2.5,1.5)}\rightarrow C_{(\rho,\rho)}^{(2.5,1.5)}$, which are GRC of $\rho$ or order $(2.25,3)$ and $(2.5,1.5)$ respectively.
 
Now, GRC's $C_{\rho,\rho}^{(\a,\b)}$ and $C_{\widehat{\rho},\widehat{\rho}}^{(\a,\b)}$ of order $(\a,\b)=(2.25,3)$ and $(2.5,1.5)$ are shown in Fig. \ref{Fig5}, for first row $\rho_{(0,1,1)}$, $\widehat{\rho}_{(0,1,1)}$ and for second row $\rho_{(1,1,1)}$, $\widehat{\rho}_{(1,1,1)}$ and $D_e=\f{7}{2},r_e=\f{1}{2},m_{\mu}=1,\hbar=1$. From this figure one can observe that  $C_{\widehat{\rho},\widehat{\rho}}^{(\a,\b)}$ is monotone and approaches to  $C_{\rho,\rho}^{(\a,\b)}$. Similarly, GRC of $\ds\widehat{\psi}_{0,0,0}({\bf r},\lam)$ with respect to $\lam$ of order $(\a,\b)=(8.5,3.5)$ is shown in Fig. \ref{Fig6} and of order $(\a,\b)=(2.25,3.5)$ is shown in Fig. \ref{Fig7} of CO, NO, N$_2$, CH, H$_2$, and ScH. On the other hand, SRC of diatomic molecules (CO, NO, N$_2$, CH, H$_2$, and ScH) of the state $\ds\widehat{\psi}_{0,0,0}({\bf r},\lam)$ with respect to $\lam$ is shown in Fig. \ref{Fig8} for $\a=2.5$ and shown in Fig. \ref{Fig9} for $\a=1.75$.

Now, one can say that, GRC and SRC are monotone and bounded functions of $|\lam|$ for all admissible values of $D_e,r_e,m_{\mu}, (n,\ell,m)$. Moreover, the R\'enyi entropy becomes negative after some values of $n$ due to irrational value of $L$. It is found that, if $C_{(\rho,\widehat{\rho}),n,\ell,m}^{(\a,\b)}$ increases, then $C_{(\widehat{\rho},\rho),n,\ell,m}^{(\a,\b)}$ decreases and vice-versa but $C_{(\rho,\widehat{\rho}),n,\ell,m}^{(\a,\b)}\ne \left[C_{(\widehat{\rho},\rho),n,\ell,m}^{(\a,\b)}\right]^{-1}$. The RCR in Eqs. (\ref{RCR.0}) and (\ref{RCR.n}) reduce to GRC in Eq. (\ref{GRC.rho}); GRC in Eqs. (\ref{GRC.rho.iso.0}) and (\ref{GRC.rho.iso.n}) reduce to GRC in Eq. (\ref{GRC.rho}) (see Fig. \ref{Fig5}) for large $|\lam|$.

Therefore, in the limiting case $|\lam|\rightarrow\infty$, the family isospectral potentials match with pseudoharomic oscillator in respect of R\'enyi entropy, RCR, GRC, SRC but the energy spacing same for all $\lam$. Therefore, we can say that, pseudoharomic oscillator and its isospectral potentials describe motion of diatomic molecules for large $|\lam|$. Energy difference bewteen of two states of pairs $(\rho_{n+1,\ell,m},\rho_{n,\ell,m})$, $(\rho_{n+1,\ell,m},\widehat{\rho}_{n,\ell,m})$, $(\widehat{\rho}_{n+1,\ell,m},\rho_{n,\ell,m})$ and $(\widehat{\rho}_{n+1,\ell,m},\widehat{\rho}_{n,\ell,m})$ are same $2\hbar\sqrt{\f{2D_e}{\mu r_e^2}}$ for all $n$, which does not match for diatomic molecule \cite{pho.import,spacing}. For any pair one can define RCR using definition (\ref{newcomplexity}) but cannot define GRC, SRC and LMC. Therefore, RCR is important to compare structure of objects which have same energy spacing. The energy spacing of these pairs for CO, NO, N$_2$, CH, H$_2$ and ScH diatomic molecules are \cite{pshop6.kds.oyewumi.jmc2012, DN.ijqc.pho} $0.203796~eV$, $0.164915~eV$,
$0.218245~eV$, $0.336462~eV$, $0.756658~eV$ and $0.155542~eV$ respectively. For non-central pseudoharmonic oscillator potential the energy spacing is not a constant \cite{DN.ijqc.pho}. In this paper potential is central, therefore, RCR's of these pairs are independent of $m$, if $\a=\b$. Now, we define RCR of them of order $(\a,\b)=(2.5,2.5)$ for $\lam=2.5$, $\ell=0$ and they are shown in Figs. \ref{Fig10} and \ref{Fig11} for different $n$. From these figures we see that RCR's are oscillate for $n$ and for large $n$ oscillation lengths decrease.

\section{Conclusion}\label{conclusion}
In conclusion, the connection between generalized R\'enyi complexity and R\'enyi complexity ratio has been established. RCR is the extension of GRC and it might be explored statistical complexities (SRC and LMC) as a particular case of RCR, depends on its order. The GRC is a product of two global information of a density function which are used in entropic uncertainty relations based on R\'enyi entropy. The GRC, SRC and LMC are interesting field of quantum chemistry for atomic structure. The RCR has been defined as a product of two global information of two density functions. Detailed mathematical characterizations of the properties of RCR have been presented. Localization property of several density functions and five theorems of near continuous property of RCR have been proved by Lebesgue measure. All these theorems would be helpful for understanding the R\'enyi continuity bound. As an example, all properties of RCR are verified for solutions of pseudoharmonic oscillator and a family of isospectral potentials. The energy levels and energy spacing in these quantum systems are same but the corresponding wave functions are different. The exact forms of R\'enyi entropy, RCR and GRC have been obtained for positive integral order and for some non-integral orders, all such measurable quantities have been calculated numerically for CO, NO, N$_2$, CH, H$_2$ and ScH. The R\'enyi entropy became negative for excited states with irrational value of $L$ for these molecules. Due to negative R\'enyi entropy, the RCR became zero for excited states with quantum number $n$. In addition, it has been found that, if $C_{(\rho,\widehat{\rho})}^{(\a,\b)}$ increases, then $C_{(\widehat{\rho},\rho)}^{(\a,\b)}$ decreases and vice-versa, but $C_{(\rho,\widehat{\rho})}^{(\a,\b)}\ne \left[C_{(\widehat{\rho},\rho)}^{(\a,\b)}\right]^{-1}$. In the limiting case $(|\lam|\rightarrow\infty)$, the R\'enyi complexity ratios $\left(C_{(\widehat{\rho},\widehat{\rho})}^{(\a,\b)},~C_{(\rho,\widehat{\rho})}^{(\a,\b)}~\&~C_{(\widehat{\rho},\rho)}^{(\a,\b)}\right)$ reduce to the generalized R\'enyi complexity $\left(C_{(\rho,\rho)}^{(\a,\b)}\right)$. The motion of diatomic molecules can be describe a family of isospectral potentials for large $|\lam|$, which agree with pseudoharomic oscillator potential. Using the definition of RCR one can compare structure of objects which have same enegy spacing. The RCR comparison for structure of objects will be very easier for central potential. Moreover, majorization effect on RCR is defined in Eq. (\ref{majo.RCR}), which is a very important property of RCR.  
%\section*{Availability of data}
%The data that supports the findings of this study are available within the article.
%\section*{Conflict of interest}
%The author has no any conflict of interest.
%\section*{ORCID}
%\noindent Debraj Nath: https://orcid.org/0000-0001-9937-7032
%\section*{References}
 

\begin{thebibliography}{99}
	\bibitem{Shannon} a) Shannon, C.E., Bell Syst. Tech. J. {\bf 27}, 379, 1948.
	b) Shannon C.E., Bell Syst. Tech. J. {\bf 27}, 623, 1948.
	\bibitem{Renyi} R\'enyi A., {\it Probability Theory}, North Holland, Amsterdam, 1970.
	\bibitem{GEUR} Portesi M., Plastino A., Physica A {\bf 225}, 412, 1996.
	\bibitem{BBM} Bialynicki-Birula I., Phys. Rev. A {\bf 74}, 052101, 2006.
	\bibitem{Renyi.stat} a) Beck C., Schl\"ogl F., {\it Thermodynamics of Chaotic Systems}, Cambridge University, Press, Cambridge, 1995.
	b) \.Zyczkowski K., Open Sys. \& Information Dyn. {\bf 10}, 297, 2003.
	\bibitem{Renyi.stat2} a) Parvan A.S., Bir\'o T.S., Phys. Lett. A {\bf 374}, 1951, 2010.
	b) Baez J.C., arXiv:1102.2098 v3.
	\bibitem{stat.complexity} Sen K.D., {\it Statistical Complexities: Application to Electronic Structure}, (Berlin: Springer 2012).
	\bibitem{Renyi.jsd} Puertas-Centeno D., Toranzo I.V., Dehesa J.S., Eur. Phys. J. Special Topics {\bf 227}, 345, 2018.
	\bibitem{Renyi.infor} G\"uhne O., Lewenstein M., Phys. Rev. A {\bf 70}, 022316, 2004.
	\bibitem{Renyi.statis} Lenzi E.K., Mendes R.S., Silva L.R. da, Physica A {\bf 280}, 337, 2000.
	\bibitem{Renyi.image} Shitong W., Chung F.L., Patt. Recognition Lett. {\bf 26}, 2309, 2005.
	\bibitem{Renyi.comp} Renner R., Wolf S., {\it International Symposium on Information Theory}, ISIT 2004. Proceedings, Chicago, IL, 2004, pp. 233, doi: 10.1109/ISIT.2004.1365269.
	\bibitem{pipek} Varga I., Pipek J., Phys. Rev. E {\bf 68}, 026202, 2003.
	\bibitem{s.liu1997} a) Liu S., Parr R.G., Phys. Rev. A {\bf 55}, 1792, 1997.
	b) Liu S., Nagy \'A., Parr R.G., Phys. Rev. A {\bf 59}, 1131, 1999. 
	c) Angulo J.C., Romera E., Dehesa J.S., J. Math. Phys. {\bf 41}, 7906, 2000.
	d) Romera E., Angulo J.C., Dehesa J.S., J. Math. Phys. {\bf 42}, 2309, 2001.
	%\bibitem{dehesa.Lp.norm} Dehesa, J.S., Guerrero, A., L\'opez, J.L., S\'anchez-Moreno, P. J. Math. Chem. {\bf 52}, 283, 2014.
	%\bibitem{dehesa.Lq.norm} Toranzo, I.V., Dehesa, J.S., S\'anches-Moreno, P. J. Math. Chem. {\bf 52}, 1372, 2014.
	\bibitem{R.G.Parr.BOOK} Parr R.G., Yang W., {\it Density-Functional Theory of Atoms and Molecules} (Oxford University Press, New York, 1989).
	\bibitem{Renyi.appl} a) Jizba P., Arimitsu T., Ann. Phys. {\bf 312}, 17, 2004.
	b) Leonenko N., Pronzato L., Savani V., Ann. Stat. {\bf 36}, 2153, 2008.
	c) Jizba P., Dunningham J.A., Joo J., Ann. Phys. {\bf 355}, 87, 2015.
	d) Toranzo I.V., Puertas-Centeno D., Dehesa J.S., Physica A {\bf 462}, 1197, 2016.
	e) Aptekarev A.I., Tulyakov D.N., Toranzo I.V., Dehesa J.S., Eur. Phys. J. B {\bf 89}, 85, 2016.
	\bibitem{Renyi.appl.nagy} Nagy \'A., Romera E., Phys. Lett A {\bf 373}, 844, 2009.
	\bibitem{Renyi.appl2} Bengtsson I., \.Zyczkowski K., {\it Geometry of Quantum States: An Introduction to Quantum Entanglement}, (Cambridge University Press, Cambridge, 2006).
	\bibitem{Renyi.nagi.mom} Romera E., Nagy \'A., Phys. Lett. A {\bf 372}, 4918, 2008.
	\bibitem{disequilibrium2} L\'opez-Ruiz R., Mancini H.L., Calbet X., Phys. Lett. A {\bf 209}, 321, 1995.
	\bibitem{lmc.plasito.continuous} Anteneodo C., Plastino A.R., Phys. Lett. A {\bf 223}, 348, 1996.
	\bibitem{lmc} Calbet X., L\'opez-Ruiz R., {\it Phys. Rev. E} {\bf 63}, 066116, 2001.
	\bibitem{lmc.continuous} Catal\'an R.G., Garay J., L\'opez-Ruiz R., Phys. Rev. E {\bf 66}, 011102, 2002.
	\bibitem{Hall} Hall M.J.W., Phys. Rev. A {\bf 59}, 2602, 1999.
	\bibitem{SDL} Shiner J.S., Davison M., Landsberg P.T., Phys. Rev. E {\bf 59}, 1459, 1999.
	\bibitem{Yamano.JMP2004} Yamano T., J. Math. Phys. {\bf 45}, 1974, 2004.
	\bibitem{lmc.biophys} L\'opez-Ruiz R., Biophys. Chem. {\bf 115}, 215, 2005.	
	\bibitem{lmc.shape.jca} L\'opez-Rosa S., Angulo J.C., Antol\'in J., Physica A {\bf 388}, 2081, 2009.
	\bibitem{lmc.appl} a) Yamano T., Physica A {\bf 340}, 131, 2004.
	b) Rosso O.A., Martin M.T., Plastino A., Physica A {\bf 347}, 444, 2005.
	c) S\'anchez J.R., L\'opez-Ruiz R., Physica A {\bf 355}, 633, 2005.
	d) Micco L.D., Gonzalez C.M., Larrondo H.A, Martin M.T., Plastino A., Rosso O.A., Physica A {\bf 387}, 3373, 2008.
	\bibitem{jc.anguloCPL} Antol\'in J., L\'opez-Rosa S., Angulo J.C., Chem. Phys. Lett. {\bf 474}, 233, 2009.
	\bibitem{a.nagy.IRP} Romera E., L\'opez-Ruiz R., Sa\~nudo S., Nagy \'A., Int. Rev. Phys. {\bf 3}, 207, 2009. arXiv:0901.1752v1	
	\bibitem{a.nagy.jmp} L\'opez-Ruiz R., Nagy \'A., Romera E., Sa\~nudo J., J. Math. Phys. {\bf 50}, 123528, 2009.
	\bibitem{a.nagy.chaos} God\'o B., Nagy \'A., Chaos {\bf 22}, 023118, 2012.
	\bibitem{jc.angulo.jsd.epjd} S\'anchez-Moreno P., Angulo J.C., Dehesa J.S., Eur. Phys. J. D {\bf 68}, 212, 2014. 
	\bibitem{jsd.pla2016} Rudnicki L., Toranzo I.V., S\'anchez-Moreno P., Dehesa J.S., Phys. Lett. A {\bf 380}, 377, 2016.
	\bibitem{DN.ijmpa} Nath D., Ghosh P., Int. J. Mod. Phys. A {\bf 34}, 1950105, 2019.
	\bibitem{DN.IJQC.Renyi} Ghosh P., Nath D., Int. J. Quantum Chem. {\bf 121}, e26461, 2021.	
	\bibitem{nagy.j.stat.mec} Romera E., Sen K.D., Nagy \'A., J. Stat. Mech: Theo. Exp. P09016, 2011.
	\bibitem{FS} Sen K.D., Antol\'in J., Angulo J.C., Phys. Rev. A {\bf 76}, 032502, 2007.
	\bibitem{Renyi.nagy.FR} Romera E., Nagy \'A., Phys. Lett. A  {\bf 372}, 6823, 2008.
	\bibitem{relative.S} Sagar R.B., Guevara N.L., J. Mol. Struct.: THEOCHEM {\bf 857}, 72, 2008.
	\bibitem{relative.F} Mukherjee N., Roy A.K., Ann. Phys. {\bf 398}, 190, 2018.
	\bibitem{Renyi.nagy.relative} Nagy \'A., Romera E., Int. J. Quantum Chem. {\bf 109}, 2490, 2009.	
	\bibitem{relative.Ts} Furuichi S., IEEE Transact. Inform. Theor. {\bf 51}, 3638, 2005.
	\bibitem{relative.lmc} Borgoo A., Geerlings P., Sen K.D., Phys. Lett. A {\bf 375}, 3829, 2011.
	\bibitem{relative.lmc.rc} Bouvrie P.A., Angulo J.C., Antol\'in J., Chem. Phys. Lett. {\bf 539}, 191, 2012.
	\bibitem{pshop9} Yahya W.A., Oyewumi K.J., Sen K.D., Int. J. Quantum Chem. {\bf 115}, 1543, 2015.
	\bibitem{selfsimilarity} Carb\'o R., Leyda L., Arnau M., Int. J. Quan. Chem. {\bf 17}, 1185, 1980.	
	\bibitem{DN.ijqc.qsi} Ghosh P., Nath D., Int. J. Quantum Chem. {\bf 121}, e26517, 2021.
	
	\bibitem{pshopEx1.2} a) Weissman Y., Jotner J., Phys. Lett. A {\bf 70}, 177, 1979.
	b) Sage M., Chem. Phys. {\bf 87}, 431, 1984.
	c) Sage M., Goodisman J., Am. J. Phys. {\bf 53}, 350, 1985.
	d) Ballahausen C.J., Chem. Phys. Lett. {\bf 151}, 428, 1988.
	e) Erkoç S., Sever R., Phys. Rev. A {\bf 37}, 2687, 1988.
	f) B\"uy\"ukkiliç F., Dem\'irhan D., Chem. Phys. Lett. {\bf 166}, 272, 1990.
	\bibitem{pshop.E.Kasap} Kasap E., G\"on\"ul B., Simsek M., Chem. Phys. Lett. {\bf 172}, 499, 1990.
	\bibitem{Flugge} Fl\"ugge S., {\it Practical Quantum Mechanics}, (Springer-Verlag, Berlin, 1994).
	\bibitem{pshopEx2} a) Popov D., Int. J. Quantum Chem. {\bf 69}, 159, 1999.
	b) Popov D., J. Phys. A: Math. Gen. {\bf 34}, 5283, 2001.
	\bibitem{pshop2.R.Sever.theo.chem} Ikhdair S., Sever R., J. Mol. Struct.: Theo. Chem. {\bf 806}, 155, 2007.
	\bibitem{pshop3.R.Sever.jmc2007} Sever R., Tezcan C., Aktas M., Yesiltas \"O., J. Math. Chem. {\bf 43}, 845, 2007.
	
	\bibitem{pshop4} a) Patil S.H., Sen K.D., Phys. Lett. A {\bf 362}, 109, 2007.
	b) Ikhdair S., Sever R., Cent. Eur. J. Phys. {\bf 5}, 516, 2007.
	c) Oyewumi K.J., Akinpelu F.O., Agboola A.D., Int. J. Theo. Phys. {\bf 47}, 1039, 2008.
	d) Ikhdair S.M., Hamzavi M., Physica B {\bf 407}, 4198, 2012.
	\bibitem{pshop5.R.Sever.jmc2012} Arda A., Sever R., J. Math. Chem. {\bf 50}, 971, 2012.
	\bibitem{pshop6.kds.oyewumi.jmc2012} Oyewumi K.J., Sen K.D., J. Math. Chem. {\bf 50}, 1039, 2012. 
	\bibitem{pshop7.R.Sever.jmc2012.1973} Akcay H., Sever R., J. Math. Chem. {\bf 50}, 1973, 2012.
	\bibitem{pshop8.H.Hassanbadi} Liu G., Guo K., Hassanabadi H., Lu L., Yazarloo B.H., Physica B {\bf 415}, 92, 2013.
	\bibitem{pshop10.F.Chand.Pramana} Rani R., Bhardwaj S.B., Chand F., Pramana J. Phys. {\bf 91}, 46, 2018.	
	\bibitem{DN.ijqc.pho} Ghosh P., Nath D., Int. J. Quantum Chem. 120, e26153, 2020.
	\bibitem{SUSY} Cooper F., Khare A., Sukhatme U.P., {\it Supersymmetry in Quantum Mechanics} (World Scientific, Singapore, 2001).
	\bibitem{tsallis} Tsallis C., J. Stat. Phys. {\bf 52}, 479, 1988.
	\bibitem{impetus} Katriel J., Sen K.D., J. Comput. Appl. Math. {\bf 233}, 1399, 2010.
	\bibitem{Onicescu} Onicescu O. C., R. Acad. Sci. Paris A {\bf 263}, 25, 1966.
		\bibitem{debnath} Debnath L., Mikusinski P., {\it Introduction to Hilbert Spaces with Applications} (Elsevier, New York, 2005).
	\bibitem{majorization} R.A. Horn, C.R. Johnson, {\it Matrix Analysis}, (Cambridge University Press), 2013, New York.
	\bibitem{joe} H. Joe, The Annals of Probability {\bf 15}, 1217, (1987).
	\bibitem{majorization.open} K. Zyczkowski, Open Sys. Information Dyn. {\bf 10}, 297, (2003).
	\bibitem{majorization.BOOK} A.W. Marshall, I. Olkin, B.C. Arnold, {\it Inequalities: Theory of Majorizations and Its Applications}, Springer, New York, USA 2010.
	\bibitem{puchala} Z. Puchala, L. Rudnicki, K. \.Zyczkowski, J. Phys. A: Math. Theor. {\bf 46}, 272002, (2013).	
	%\bibitem{Renyi.nature} Hu X., Ye Z., J. Math. Phys. {\bf 47}, 023502, 2006.
	\bibitem{jsd.jmp.Renyi.bound} S\'anchez-Moreno P., Zoror S., Dehesa J.S., J. Math. Phys.{\bf 52}, 022105, 2011. 	

%	\bibitem{severJMC2009.46.1122.24.25} a) Bransden, B.H., Joachain, C.J. {\it Physics of Atoms and Molecules}, 2nd edn., chaps. X-XI (Pearson Education, Harlow, England, 2003).\\	b) Ogilvie, J.F. {\it The vibrational and Rotational Spectrometry of Diatomic molecules}, chap. IV (Academic Press, San Diago-California, USA, 1998).
	\bibitem{pho.import} Herzberg G., {\it Molecular Spectra and Molecular Structure. I. Spectra of Diatomic Molecules} (New York: Van Nostrand Reinhold) 1950.
	\bibitem{Ryzhik} Gradshteyn I.S., Ryzhik I.M., {\it Table of Integrals, Series and Products}, (Academic, New York, 1994).
	\bibitem{DN.jmc} Nath D., J. Math. Chem. {\bf 51}, 1446, 2013.
	\bibitem{DN.ijqc.iso} Ghosh P., Nath D., Int. J. Quantum Chem. 119, e25964, 2019. 
	\bibitem{Srivastava} a) Srivastava H.M., Karlsson P.W., {\it Multiple Gaussian Hypergeometric Series}, John Wiley and Sons, New York, 1985.
	b) Srivastava H.M., Niukkanen A.W., Math. Comput. Model. {\bf 37}, 245, 2003.
	\bibitem{jsd.ijqc2016} Dehesa J.S., Toranzo I.V., Puertas-Centeno D., Int. J. Quantum Chem. {\bf 117}, 48, 2016.
	\bibitem{jsd.Laguerre} S\'anchez-Moreno P., Manzano D., Dehesa J.S., J. Comput. Appl. Math. {\bf 235}, 1129, 2011.
	\bibitem{jsd.ijqc2011} S\'anches-Moreno P., Omitse J.J., Dehesa J.S., Int. J. Quantum Chem. {\bf 111}, 2283, 2011.
	\bibitem{jsd.amc} S\'anchez-Moreno P., Dehesa J.S., Zarzo A., Guerrero A., Appl. Math. Comput. {\bf 223}, 25, 2013.
	%\bibitem{Kratzer.8kds.ijc} Yahya W.A., Oyewumi K.J., Sen K.D., Indian J. Chem. {\bf 53A}, 1307, 2014.
	\bibitem{data} a) Nakamura K., et al. (Particle Data Group), J. Phys. G {\bf 37}, 075021, 2010. b) Falaye B.J., Oyewumi K.J., Sadikoglu F., J. Theor. Comt. Chem. {\bf 14}, 1550036, 2015.
	\bibitem{spacing} a) Maiz F., Phys. Scr. {\bf 96}, 105403, 2020.
	b) Fernández M., 2021 Phys. Scr. {\bf 6}, 077001, 2021.
	
	
	
\end{thebibliography}
\end{document}